\documentclass{aa}  

\usepackage{graphicx}
\usepackage{txfonts}
\begin{document}

   \title{Properties of extragalactic thick discs recovered from ultra-deep Stripe82 imaging}

   \author{C. Mart{\'i}nez-Lombilla
          \inst{1,2}
          \and
          J. H. Knapen\inst{1,2,3}
          }

   \institute{Instituto de Astrof{\'i}sica de Canarias (IAC), La Laguna, 38205, Spain\\
              \email{cml@iac.es}
         \and
             Departamento de Astrof{\'i}sica, Universidad de La Laguna (ULL), E-38200, La Laguna, Spain
         \and
            Astrophysics Research Institute, Liverpool John Moores University, IC2, Liverpool Science Park, 146 Brownlow Hill, Liverpool, L3 5RF, UK\\
             }

   \date{January 14, 2019}

% \abstract{}{}{}{}{} 
% 5 {} token are mandatory
 
  \abstract
   {Thick discs can give invaluable information on the formation and evolution history of galaxies as most, if not all, disc galaxies have a thin (classical) disc and a thick disc.}
   {We study the structure of thick discs in extraordinary depth by reaching a surface brightness limit of $\mu_{r_{\rm deep}} \sim$28.5$-$29~mag~arcsec$^{-2}$ with combined $g$,$r$,$i$ band images from the IAC Stripe 82 Legacy Project.}
   {We present the characterisation of the thick discs in a sample of five edge-on galaxies. A study of the radial and vertical surface brightness profiles is presented by comparing our data with point spread function (PSF) deconvolved models. Our method begins with an analysis of the background and masking processes. Then we consider the effects of the PSF through galaxy modelling. The galaxy disc components are fitted considering that the thin and thick discs are two stellar fluids that are gravitationally coupled in hydrostatic equilibrium.}
   {We find that effects due to the PSF are significant when low surface brightness is reached, especially in the vertical profiles, but it can be accounted for by careful modelling. The galaxy outskirts are strongly affected by the faint wings of the PSF, mainly by PSF-redistributed light from the thin disc. This is a central problem for ultra-deep imaging. The thick-disc component is required to reach satisfactory fit results in the more complex galaxies in our sample, although it is not required for all galaxies. When the PSF is ignored, the brightness of these structures may be overestimated by up to a factor of $\sim$4.
}
   {In general, our results are in good agreement with those of previous works, although we reach deeper surface brightness levels, so that the PSF effects are stronger. We obtain scale heights and mass ratios of thin and thick discs ($z_{{\rm t}}$, $z_{{\rm T}}$, and $M_{{\rm T}}/M_{{\rm t}}$), which provide excellent agreement with previous studies. Our small initial sample provides evidence for aspects of a wide variety of formation theories for the thick discs in disc galaxies.}

   \keywords{Methods: observational --
            Methods: data analysis --
            Techniques: image processing --
            Galaxies: formation --
            Galaxies: evolution --
            Galaxies: structure
            }

   \maketitle
%
%________________________________________________________________

\section{Introduction}\label{intro}
Most if not all disc galaxies have a thin (or classical) disc and a thick disc \citep{Burstein1979, Tsikoudi1979, Yoshii1982, Fuhrmann1998, Dalcanton2002, Yoachim2006, Fuhrmann2011, Snaith2014, Comeron2011, Comeron2012, Comeron2014, Comeron2017}. In most models thick discs are thought to be a necessary consequence of the disc formation process and/or of the evolution of the galaxy. However, because of their faint surface brightness ($\mu_{g} \sim$26.0 mag~arcsec$^{-2}$), the detailed study of the physical properties of thick discs has thus far been very difficult. The goal of this paper is to shed light on this problem by characterising the thick-disc component in an initial sample of five edge-on nearby spiral galaxies in order to compare the properties of their thin and thick discs.
    
The origin of thick discs is still a matter of debate \citep[][and references therein]{Yoachim2008a}. Although there are many different approaches, three main families of models have been proposed \citep[e.g.][]{Comeron2011}. The first family of models is secular. Here, the original idea was that the thick disc was caused by vertical heating and/or radial migration of stars \citep[e.g.][]{Schoenrich2009, Loebman2011}, but radial migration was later found to be less likely as a mechanism \citep[e.g.][]{Minchev2012, VeraCiro2014} and a revised scenario has been proposed where thick discs  form from the nested flares of mono-age populations \citep{Minchev2015, Martig2016}. In the second family, the disc was always thick from its beginning, as a result of a high-velocity dispersion of the interstellar medium at high redshift; the thin disc is accreted with a lower velocity dispersion \citep[e.g.][]{Elmegreen2006, Bournaud2009}. In the third class of models, interactions with satellite galaxies either due to dynamical heating \citep[e.g.][]{Quinn1993} or accretion of stars \citep[e.g.][]{Abadi2003} are responsible for the formation scenario of the thick disc. The three families of models suggest distinct outcomes depending on the physical properties of the thick disc. Thus, a proper analysis of their parameters can provide constraints on disc formation models.

Thick discs can give invaluable information about the evolution of galaxy structures. Our goal is to study their structure in unprecedented detail. However, the study of low surface brightness structures is a technical challenge from an observational point of view. In the past several years  significant advances have been made in detector technology, observational strategies, and data reduction techniques in order to fulfil the requirements of new ultra-deep imaging science. These improvements allow astrophysicists to reach surface brightnesses fainter than 28~mag~arcsec$^{-2}$ in the $r$ band. These depths have been achieved on many scales, from large imaging surveys to small galaxy samples, or even in the study of individual objects. These observational depths may be achieved either through very long exposures on small telescopes or using shorter exposures on the largest telescopes. The most remarkable works have achieved a surface brightness of $\mu_{r} \sim 29$~mag~arcsec$^{-2}$ \citep[3$\sigma$, 10$\times$10~arcsec boxes; e.g.][]{Martinez-Delgado2010, Ferrarese2012, Merritt2014, Duc2014, Koda2015, Capaccioli2015, Peters2017, Martinez-Lombilla2018}, which is equivalent to a surface brightness $\sim$1500 times fainter than the darkest sky on Earth. Only a few studies have passed beyond the $\mu_{r} = 30$~mag~arcsec$^{-2}$ frontier \citep[3$\sigma$, 10$\times$10~arcsec boxes; e.g.][]{TrujilloFliri2016, Borlaff2019}. Further progress in the field of low surface brightness science is difficult not because of the collection capability of the telescopes and their detectors, which provide deep imaging and good photon statistics, but rather because of systematic errors. It is therefore crucial to perform very careful data reduction and treatment of both sky and object images \citep[for a more detailed description, see the review by][]{Knapen2017a}. For this purpose, we have used very deep data that reach surface brightness limits of 28.5$-$29~mag~arcsec$^{-2}$ (3$\sigma$, 10$\times$10~arcsec boxes) from the combined $g$-, $r$-, and $i$ -band images from the IAC Stripe 82 Legacy Project \citep{FliriTrujillo2016}. 

The light of a source is affected by the telescope, the instrument, and the detector, as well as by the atmosphere. This results, among other effects, in an increase of the scattered light. The point spread function (PSF) characterises how the light of a source is affected and measures the extent of that scattered light. How to account for the PSF effect on astronomical objects and the importance of this process is well documented \citep[e.g.][]{Michard2002, Slater2009, Sandin2015, TrujilloFliri2016}. However, the correction process is still challenging. The shape of the PSF may vary widely depending on the atmospheric conditions and the instrument used, and is generally time-variable on a timescale that can be as short as minutes. The projected surface brightness structure of a source is convolved with the PSF to make up the observed form. If not corrected for, the scattered light therefore adds a systematic component to the observed intensities of our low surface brightness objects. This effect extents to large angular radii, so that although the PSF rapidly becomes faint with increasing radius, the integrated amount of light in its faint extended wings can contribute significantly to observed structures. In the case of extended objects such as galaxies, \cite{Sandin2014,Sandin2015} found that the scattered light of such sources (if not accounted for) can in principle be misinterpreted as a bright stellar halo or a thick disc. Other works have shown that the properties of stellar haloes and the faint outskirts of galactic discs can indeed be significantly influenced by the wings of the stellar PSFs \citep[e.g.][]{Zibetti2004a, deJong2008, Sandin2014, TrujilloFliri2016, Peters2017, Comeron2017}. In general, the effect becomes stronger the sharper the light distribution of the source \citep{Trujillo2001a, Trujillo2001b}. Thus, a proper consideration of the scattered light distribution is crucial, especially in light of upcoming all-sky deep surveys such as the LSST \citep[Large Synoptic Survey Telescope;][]{Gressler2014}. 

In this work, we characterise the thick discs in five edge-on galaxies with a wide range of mass. A study of the radial and vertical surface brightness profiles is presented by comparing our data with model predictions. We find that a thick-disc component is required to reach satisfactory fit results, especially in intermediate- to high-mass galaxies, but its brightness is overestimated if the PSF is ignored. Thus, effects due to the PSF are significant when a lower surface brightness is reached, especially in vertical profiles, but this can be accounted for by careful modelling. We show that galaxy outskirts are strongly affected by the faint wings of the PSF, mainly by light from the thin disc that is redistributed by the PSF.

\section{IAC Stripe 82 Legacy Project} \label{chap3/sec2:IAC_Stripe 82}

Our data are taken from the publicly available IAC Stripe 82 Legacy Project\footnote{Public data release: http://www.iac.es/proyecto/stripe82} \citep[-50$^{\circ}$$ \leq$\,RA\,$\leq$\,+60$^{\circ}$ and -1.25$^{\circ}$\,$\leq$\,Dec.\,$\leq$\,+1.25$^{\circ}$;][]{FliriTrujillo2016}. This project produced optimal co-adds of the SDSS Stripe 82 dataset after reprocessing of all images. The main goal of the IAC Stripe 82 Legacy Project is to maintain the low surface brightness features on all spatial and intensity scales using a non-aggressive sky-subtraction strategy. Thus, the IAC team in particular preserves the characteristics of the background during the reduction, that is, the sky and the diffuse light. The IAC Stripe 82 Legacy Project reduction allows us to reach a surface brightness limit of $\sim$28.5\,mag arcsec$^{-2}$ in the \textit{r} band (3\,$\sigma$, 10$\times$10\,arcsec$^{2}$), which improves by 1.7-2.0 mag upon the single-epoch SDSS images. Thus, it is possible to trace surface brightness profiles down to $\sim$~30\,\textit{r}-mag arcsec$^{-2}$ \citep{Peters2017}, a depth that before has been reached by only a few studies \citep[e.g.][]{Zibetti2004a, Zibetti2004b, Jablonka2010, TrujilloBakos2013, TrujilloFliri2016}. Our images are in the deepest band, a deep \textit{r} band co-add ($r_{\rm deep}$), which is the linear average of the \textit{g, r,} and \textit{i} co-adds.

In order to determine whether the faint wings of the PSF can add artificial light to faint components in the outskirts of galaxies, we need to construct our own PSF to model the galaxies. To estimate this effect in the Stripe 82 data, the IAC Stripe 82 Legacy Project provides ultra-deep and very extended PSFs that extend to 8 arcmin in radius and to over 20 mag in dynamic range for any particular field \citep[details in][]{FliriTrujillo2016}. We use the ultra-deep extended PSFs in the $r_{\rm deep}$ band of our regions of interest (FWHM$_{r_{\rm deep}}= 1.08$~arcsec), obtained by combining the wings of very bright saturated stars from the whole Stripe 82 area (i.e., the outer part of the extended PSF) and \verb|PSFEx|-generated PSFs with high signal-to-noise ratio (S/N) of a particular field (i.e. the inner part of the extended PSF).

%The IAC team calculated the PSFs using the \verb|PSFEx| software which selects point sources (between 300 and 2000 depending on the location of the field and the filter) using the half-light radius and luminosity information included in the Source Extractor \citep{Bertin1996} catalogues. A robust mean PSF for a particular field is obtained by the normalization and combination of all those point sources. Ultra-deep extended PSFs in all filters (the five SDSS bands plus the $r_{\rm deep}$) were obtained by combining the wings of very bright saturated stars from the whole Stripe 82 area (i.e., the outer part of the extended PSF) and \verb|PSFEx|-generated high-S/N (signal to noise ratio) PSFs of a particular field (i.e., the inner part of the extended PSF). The profiles of saturated and non-saturated PSFs are matched to align the flux scaling. Finally, both PSFs were combined to build an ultra-deep extended PSF for each field and each band. We show an example of an ultra-deep extended PSF in the $r_{\rm deep}$ band in Fig.~\ref{figure//chap3/sect2:PSF_Stripe}.

%As expected, the Stripe 82 ultra-deep PSF is narrower in the redder bands, with FWHM values in \textit{r}, \textit{i} and \textit{z} below 1.1 arcsec: $u_{\rm FWHM} = 1.31$~arcsec, $g_{\rm FWHM} = 1.24$~arcsec, $r_{\rm FWHM} =$~$1.10$~arcsec, $i_{\rm FWHM} = 1.02$~arcsec, $z_{\rm FWHM} = 1.04$~arcsec, and $r_{\rm deep, FWHM} = 1.08$~arcsec. However, there is a strong dependence of the median width of the PSF on the declination of the field, while the dependence on right ascension is very weak.

\section{Selection criteria and final sample} \label{chap3/sect3:TargetSelection}

%\begin{figure*}
%\begin{center}
%\includegraphics[width=157mm]{56sample}
%\includegraphics[width=157mm]{7sample}
%\caption[ ]{Top: Shortlist of 56 galaxies after performing the first selection using the HyperLeda database. We show an optical colour image from the SDSS Data Release 12 Navigate Tool of each galaxy. The name and the galaxy coordinates (right ascension and declination) are located at the top of each panel. Bottom: shortlist of 5 selected galaxies after visual inspection of the sample of 56 pre-selected targets; the primary reason why we reject a galaxy is shown.}
%\label{figure//chap3/sect3:7sel}
%\end{center}
%\end{figure*}

\begin{table*}
\begin{center}
\footnotesize
\caption{List of the 51 rejected galaxies from the shortlist of 56 galaxies after performing the first selection using the HyperLeda database. The name and the primary reason why we rejected the galaxy from the final data sample are indicated in each column.}

\begin{tabular}{l|r}
Galaxy ID & Rejection Reason \\
\hline
2MASX~J2231+0026 & No distance \\
IC~0330 & Not edge-on \\
IC~1756 & Foreground objects \\
NGC~0585 & Dust lane \\
NGC~0955 & Dust lane \\
NGC~7648 & No disc galaxy \\
PGC~011579 & Not edge-on \\
PGC~011777 & No distance \\
PGC~012721 & No distance \\
PGC~066048 & Foreground objects \\
PGC~072794 & Not edge-on \\
PGC~091784 & No distance \\
PGC~092868 & Foreground objects \\
PGC~1149944 & No distance \\
PGC~1157026 & Not edge-on \\
PGC~1171976 & Interaction\\
PGC~173375 & No distance \\
\end{tabular}
\hfill
\begin{tabular}{l|r}
Galaxy ID & Rejection Reason \\
\hline
PGC~191337 & Not edge-on \\
PGC~3107779 & No distance \\
SDSS~J0242-0053 & No distance \\
UGC~00507 & Dust lane \\
UGC~00651 & Not edge-on \\
UGC~00734 & Not edge-on \\
UGC~00847 & Very thin disc \\
UGC~00866 & Not edge-on \\
UGC~00961 & No distance \\
UGC~01116 & Not edge-on \\
UGC~01120 & Not edge-on \\
UGC~01123 & Not edge-on \\
UGC~01296 & Not edge-on \\
UGC~01588 & Not edge-on \\
UGC~01911 & Foreground objects \\
UGC~01934 & Not edge-on \\
UGC~02319 & Not edge-on \\
\end{tabular}
\hfill
\begin{tabular}{l|r}
Galaxy ID & Rejection Reason \\
\hline
UGC~02404 & Foreground objects \\
UGC~02522 & Not edge-on \\
UGC~02523 & Not edge-on \\
UGC~02584 & No distance \\
UGC~02587 & Clipped image\\ 
UGC~02628 & Not edge-on \\
UGC~02645 & Foreground objects \\
UGC~02699 & No distance \\
UGC~11612 & Foreground objects \\
UGC~11631 & Warp \\
UGC~11645 & Foreground objects \\
UGC~11724 & Not edge-on \\
UGC~11859 & Very thin disc \\
UGC~11982 & Not edge-on \\
UGC~12183 & Not edge-on \\
UGC~12348 & Not edge-on \\
UGC~12865 & Warp \\
\end{tabular}
\label{table//chap3/sect3:7sel}
\end{center}

\tablefoot{There was a problem with UGC~02587 because its images are clipped, so that the galaxy falls on one of the edges of the field. Thus, half of the galaxy is in one image and the other half in another. In addition, UGC~02587 has a strong dust lane and several foreground objects, so that even with a perfect match between frames by using astronomical software such as SWarp \citep{Bertin2002}, the image would not be useful for scientific purposes. We reject this galaxy for these reasons, but also because it requires different background treatment in each of the frames, which could affect the consistency of the surface brightness measurements at very faint levels, especially in the boundary region of both images.}

\end{table*}

Thick discs are low surface brightness structures that have been detected by different observational approaches. In particular, their study using integrated light of edge-on galaxies has been performed by \cite{Yoachim2006} and \cite{Comeron2011, Comeron2012, Comeron2017}, for instance. Thus, we need to search for highly inclined disc galaxies with available deep imaging in a multitude of bands and with a spatial resolution that is high enough to resolve these structures. We now describe the selection criteria required to obtain a reliable and unbiased sample of nearby highly inclined galaxies that are imaged in the SDSS \textit{g, r,} and \textit{i} bands.

In the first step we used the HyperLeda\footnote{http://leda.univ-lyon1.fr/} online catalogue \citep{Makarov2014} to select all highly inclined galaxies (inclination $\geq 70^{\circ}$ or log$_{\rm{r25}} \geq 0.477$) that fall within the IAC Stripe 82 Legacy Project area ($\sim$ 275\,deg$^{2}$,  $-$50$^{\circ} \leq$\,RA\,$\leq$\,$+$60$^{\circ}$ and $-$1.25$^{\circ}$\,$\leq$\,Dec.\,$\leq$\,$+$1.25$^{\circ}$) and have large apparent sizes ($d_{\rm{d25}}\geq$1 arcmin or log$_{\rm{d25}}\geq 1$) in order to have enough spatial resolution in the discs. At this stage, we had a pre-selection of 56 galaxies. The next step was to make a second-order selection by visual inspection of the SDSS Data Release 12 colour images. Table~\ref{table//chap3/sect3:7sel} shows the list of 51 rejected galaxies from that pre-selected sample as well as the primary reason for rejection for all of them.

\begin{figure}
\begin{center}
\includegraphics[width=80mm]{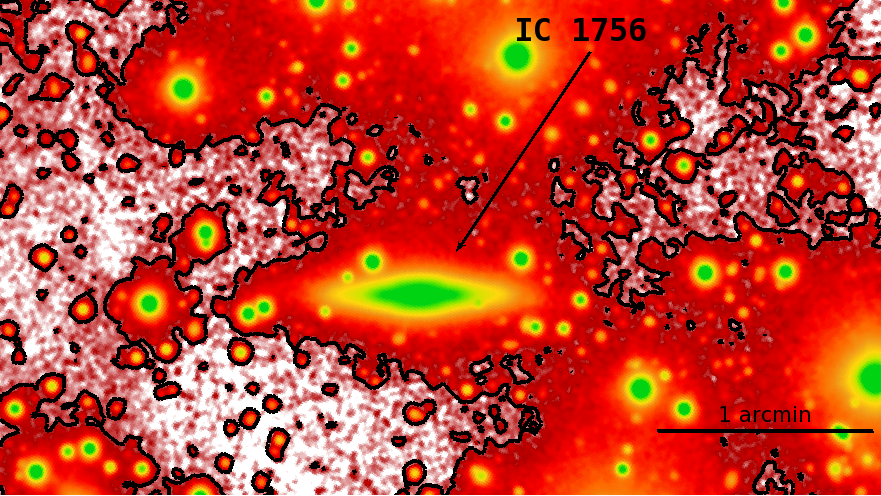}
\caption[]{Detail of the area around galaxy IC~1756. The brightness of foreground stars adds light to the faint galaxy structures, as the isocontour at 28~mag~arcsec$^{-2}$ shows. The high density of objects in the region also affects the surface brightness properties of our target, and therefore many masking areas would be needed (see Sect.~\ref{chap3/sect4:Masks}) in order to keep only the galaxy light in our study. For this reason, this galaxy was deemed unsuitable for further analysis here.}
%
% You may add a label to this figure for future reference.
%
\label{figure//chap3/sect3:crowdField}
\end{center}
\end{figure}

Morphological features such as warps or possible interacting satellites are undesirable. We also rejected galaxies with very bright foreground objects in the field because they can affect the measurement results. Crowded areas are not recommended, especially if the area of the thick disc is affected. To illustrate this, Fig.~\ref{figure//chap3/sect3:crowdField} shows the bright stars around IC~1756 in detail. It is clear that the extremely crowded region adds undesired light to the low surface brightness galaxy structures.

\begin{figure*}
\begin{center}
\includegraphics[width=.3\linewidth]{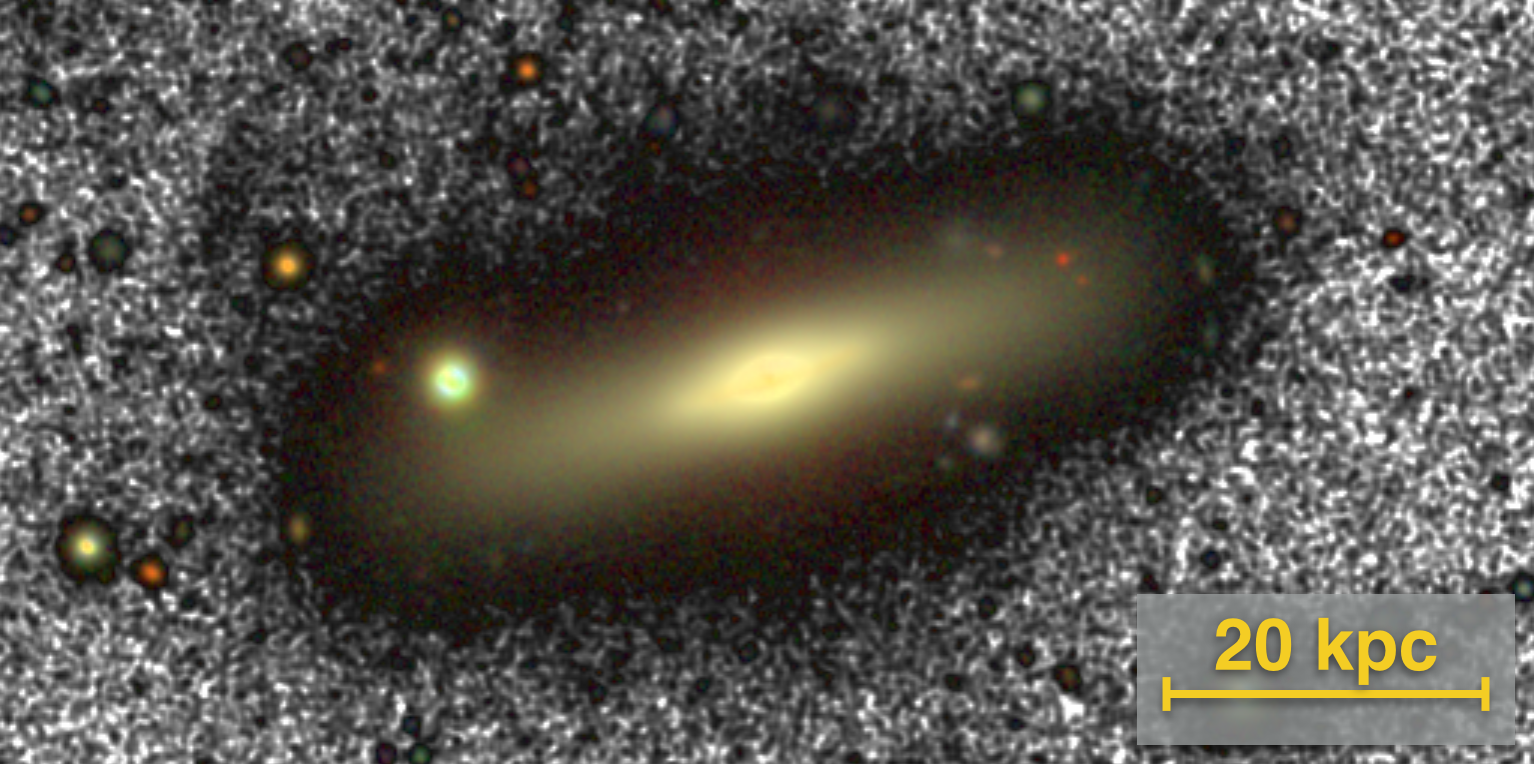}  
\includegraphics[width=.28\linewidth]{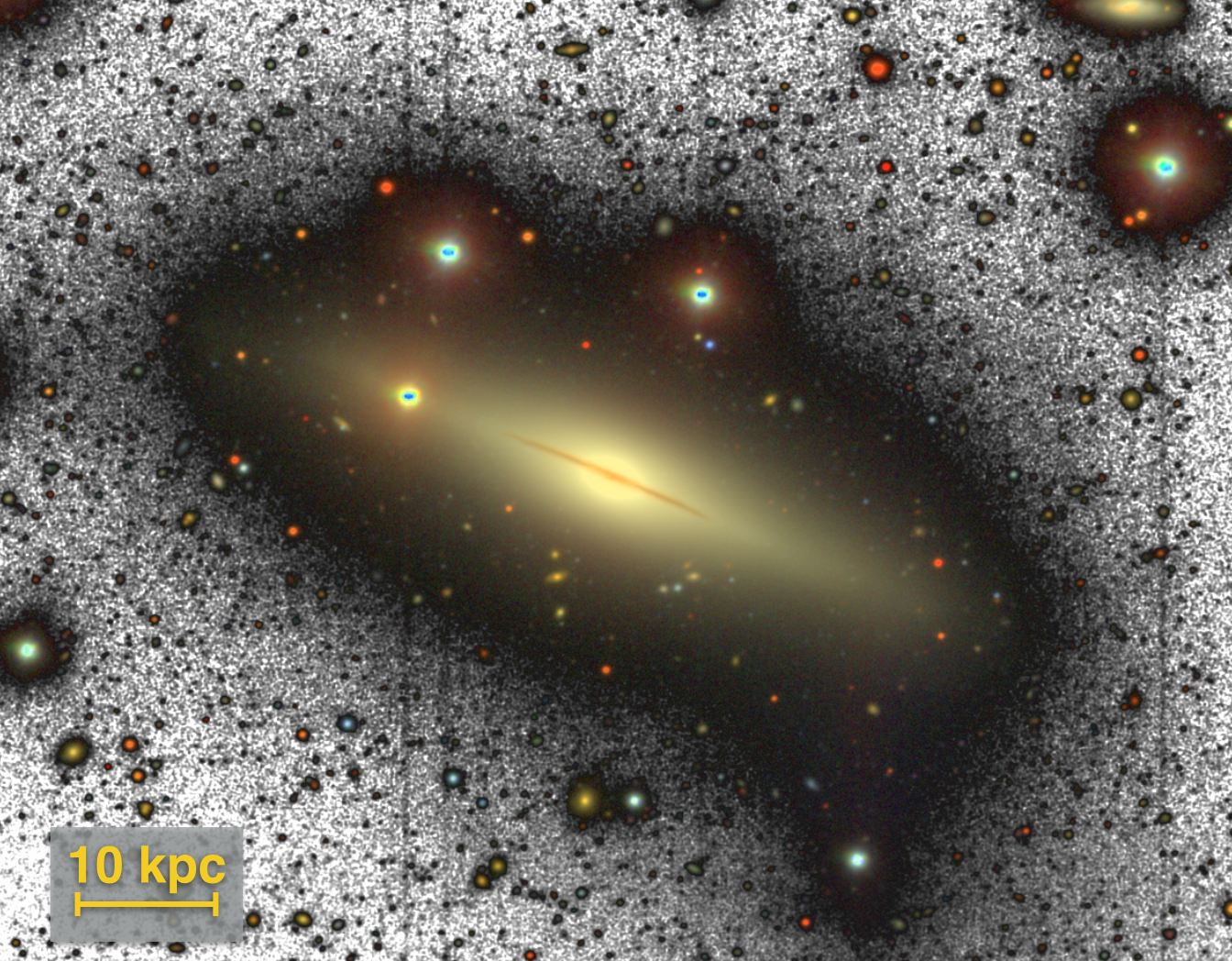}  
\includegraphics[width=.3\linewidth]{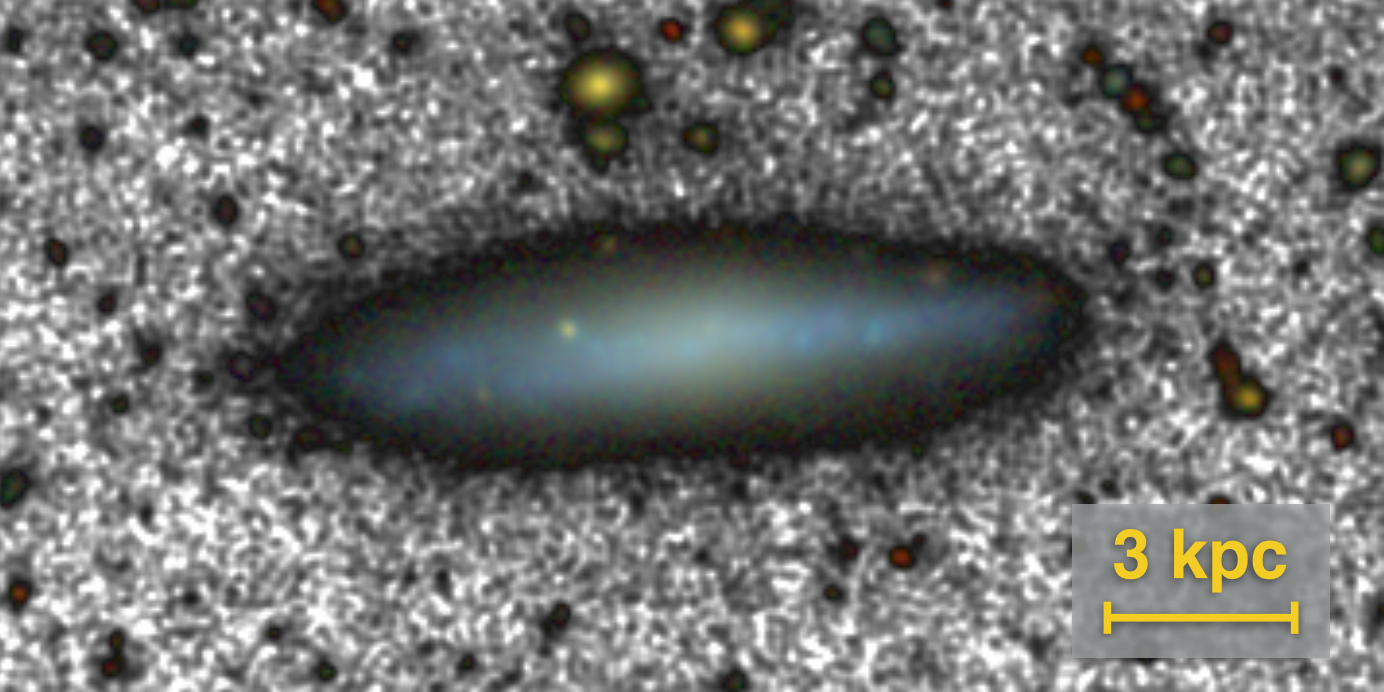}  
\includegraphics[width=.213\linewidth]{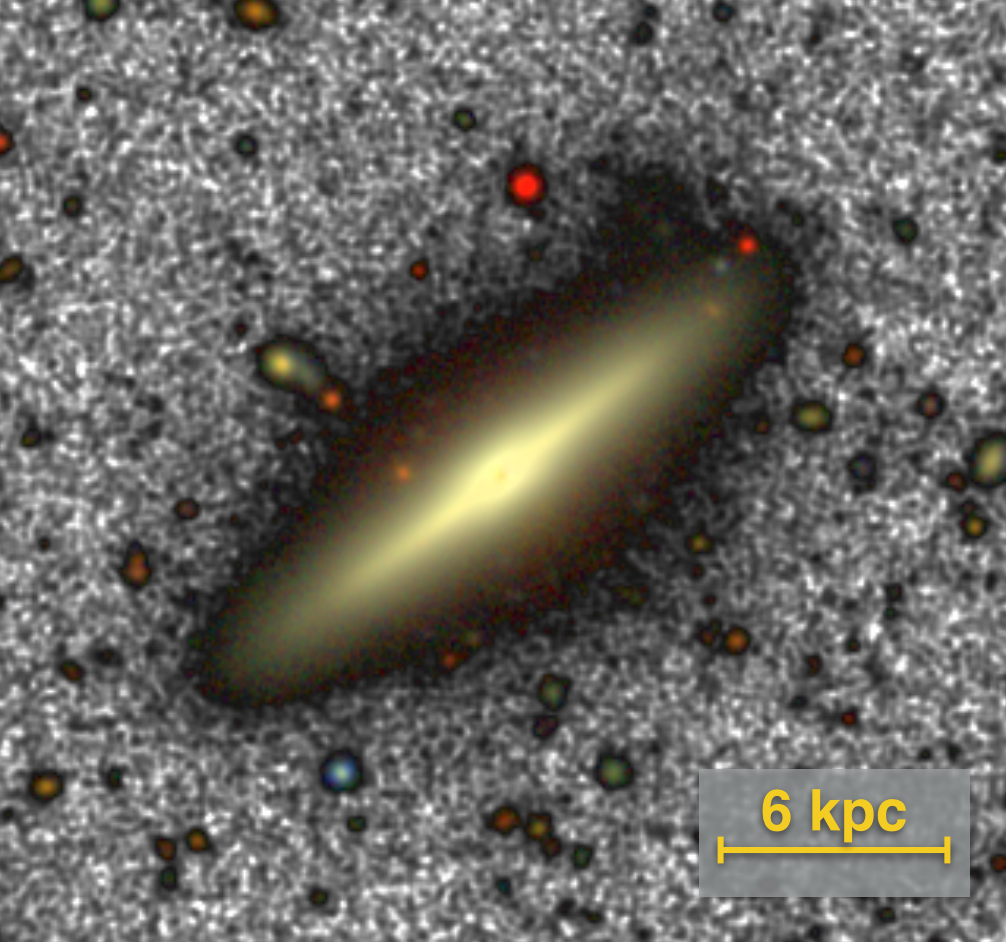}  
\includegraphics[width=.3\linewidth]{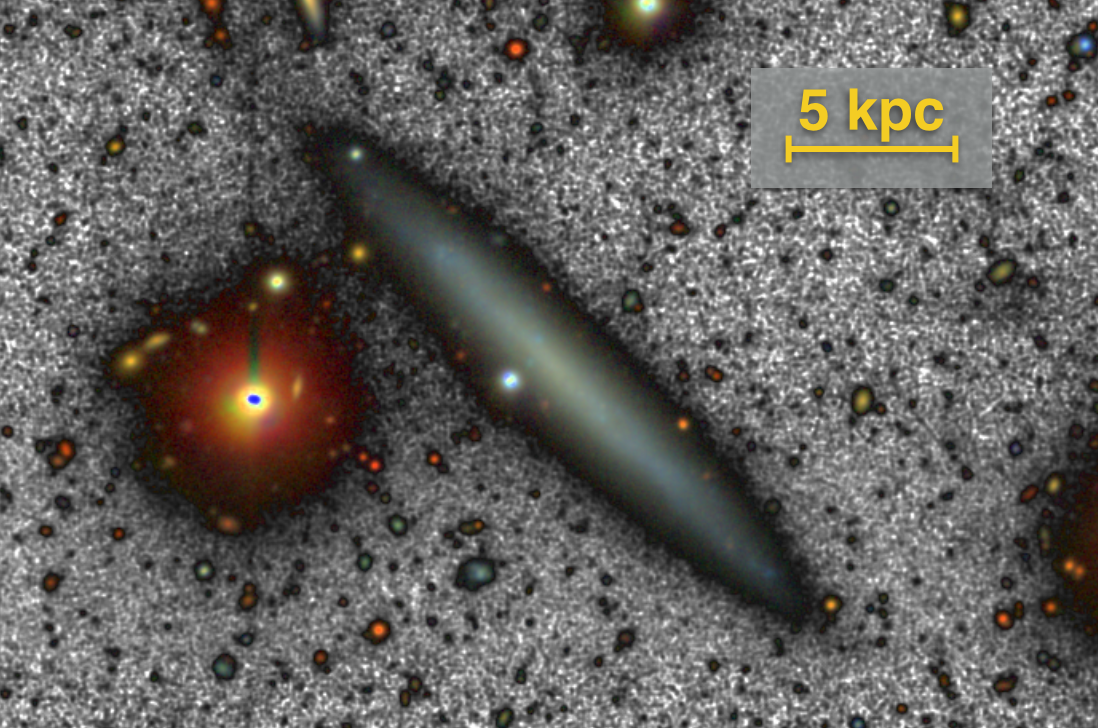} 
\end{center}
\protect\caption[ ]{SDSS optical colour images of the final galaxy sample (roughly equivalent to an \textit{r}-band image), shown overlaid on the $r_{\rm deep}$ IAC Stripe 82 Legacy Project images in order to illustrate the data quality. From top to botton and from left to right: NGC~0429, NGC~1032, UGC~00931, UGC~01040, and UGC~01839.}
\label{figure//chap3/Ind_galaxies} 
\end{figure*}

In order to have discs that are sufficiently visible to derive reliable radial and vertical surface brightness profiles, we rejected galaxies with insufficiently high inclination or non-disc galaxies, as well as galaxies with very strong dust lanes. The latter is particularly noteworthy when the dust lanes are irregular (i.e. not just a straight line) and/or when they are thick and cover most of the visible disc. In the particular case of NGC~1032 (see Fig.~\ref{figure//chap3/Ind_galaxies}), the dust lane is extremely thin and lies over the very central part of the disc, so that we are able to study the dust-free regions of the thin- and thick-disc structures. This galaxy is especially interesting as it is the only one in our sample that was studied in previous papers on thick discs using S$^{4}$G data \citep{Comeron2011, Comeron2012, Comeron2014, Comeron2017}.

As mentioned earlier, we require galaxies of a large apparent size to resolve the disc structure. For the same reason, we need stellar discs with a width that allows us to distinguish thick-disc structures.

As our study is focused on the analysis of disc components, we need to know the distance to the object in order to obtain the physical size and the extent of both the thin and thick disc. Thus, the final selection criterion is the availability of distance measurements in the on-line databases NASA$/$IPAC Extragalactic Database\footnote{http://ned.ipac.caltech.edu/} \citep[NED:][]{Helou1991} and SIMBAD\footnote{http://simbad.u-strasbg.fr/simbad/} \citep{Wenger2000}.

\begin{table*}
\begin{center}
\footnotesize
\caption{Final data sample properties and physical parameters.} 
\label{table//chap3/sect3:finalSample} 
{
\begin{tabular}{c||c|c|c|c|c|c|c|c|c}
\hline 
Galaxy & Morph. & D & Incl. & M$_{abs}$ & $v_{rot}$  & $M_{\rm{dyn}}$ & M$_{\star}$ & $d_{25}$ & $\mu_{r_{\rm deep}}$ \\  
 ID & type & [Mpc] & [deg.] & [B-mag] & [km/s] & [M$_{\odot}$] & [M$_{\odot}$] & [arcmin] & ~mag~arcsec$^{-2}$ \\
\hline  
NGC~0429    & S0-b?                 & 93.5    & 90.0   & -20.29 $\pm$ 0.353     & -             & -               & 4.4-8.3$\times$10$^{10}$  & 1.51 $\pm$ 0.11 & 28.5 $\pm$ 0.1 \\ 

NGC~1032    & S0$^{-}$sp / E(d)7    & 38.7    & 88.1   & -20.75 $\pm$ 0.43    & 261.0         & 3.3$\times$10$^{11}$  & -                     &  3.55 $\pm$ 0.11 & 29.7 $\pm$ 0.3 \\ 
 
UGC~00931   & Sc                & 31.9    & 78.1  & -17.46 $\pm$ 0.678    & 50.7 $\pm$ 2.1  & 3.1$\times$10$^{9}$   & -                     &  1.02 $\pm$ 0.11 & 29.3 $\pm$ 0.2 \\ 
 
UGC~01040   & S0-a?             & 44.1    & 90.0   & -19.62 $\pm$ 0.37    & -             & -                 & 1.3-2.8$\times$10$^{10}$    &  1.17 $\pm$ 0.12 & 29.0 $\pm$ 0.2 \\ 
  
UGC~01839   &  Sd\underline{m}:sp   & 24.7    & 90.0   & -17.40 $\pm$ 0.517   & 65.1          & 5.9$\times$10$^{9}$   & -                     &   1.20 $\pm$ 0.12 & 28.0 $\pm$ 0.1 \\

\hline 
\end{tabular}
}
\end{center}

\tablefoot{The columns represent (1) the galaxy ID; (2) morphological type in the de Vaucouleurs system from the HyperLeda database \citep{Makarov2014} for NGC~0429, UGC~01040, and UGC~00931 and morphological type from \cite{Buta2015} for NGC~1032 and UGC~01839; (3) distance in Mpc from NED \citep[][]{Helou1991}; (4) inclination to the line of sight from the HyperLeda database; (5) absolute B-band magnitude (HyperLeda database); (6) maximum rotational velocity corrected for inclination (HyperLeda database); (7) dynamical mass (HyperLeda database); (8) stellar mass from \cite{Chang2015}; (9) diameter in arcmin at the isophotal level 25~mag~arcsec$^{-2}$ in the B band corrected for galactic extinction (HyperLeda database); (10) surface brightness limit of the IAC Stripe 82 Legacy Project $r_{\rm deep}$ band images, obtained as the equivalent to a 3\,$\sigma$ fluctuation (compared to the sky noise) in square boxes of 10$\times$10\,arcsec$^{2}$.}
\end{table*}

The final sample of galaxies consists of three massive galaxies with complex structures (NGC~0429, NGC~1032, and UGC~01040) and two low-mass galaxies with a simple morphology (UGC~01839 and UGC~00931) from a total initial sample of 56 galaxies. Table~\ref{table//chap3/sect3:finalSample} summarises the general information and physical parameters of the five galaxies in the final sample. The Hubble type was taken from the de Vaucouleurs system from the HyperLeda database \citep[for NGC~0429, UGC~01040 and UGC~00931; ][]{Makarov2014} or from the \textit{Spitzer} Survey of Stellar Structure in Galaxies (S$^{4}$G) morphology classification of \cite{Buta2015} in the case of NGC~1032 and UGC~01839. The distance was obtained from the average of the most recent values (since 2010) in NED \citep[][]{Helou1991}. The other properties were all taken from the HyperLeda database \citep{Makarov2014}. When the maximum rotational velocity was not available in Hyperleda, we took the values of the stellar mass from \cite{Chang2015} instead. Fig.~\ref{figure//chap3/Ind_galaxies} shows the SDSS optical colour images of the final galaxy sample, shown overlaid on the $r_{\rm deep}$ IAC Stripe 82 Legacy Project images in order to illustrate the data quality.

\section{Method}

\subsection{Masking process} \label{chap3/sect4:Masks}

Ultra-deep imaging requires an accurate background treatment to remove the sky and foreground objects \citep[see, e.g. ][]{TrujilloFliri2016}. The goal of the masking process is to avoid any undesirable flux from fore- and background objects over the target in order to only keep the flux from the source we wish to analyse. However, this is difficult because we reach surface brightness levels of $\sim$ 29-30 r-mag~arcsec$^{-2}$. At these depths, even the light from bright stars far outside the frame can contribute to the galaxy flux. To address this problem, we used a semi-automatic code developed by \citet[][see details in their Section~3.1]{Martinez-Lombilla2018}. It was optimised for low surface brightness data, allowing us to build a mask of each image covering every emission source except for the galaxy to be analysed.

\subsection{Sky subtraction and background treatment}\label{chap3/sect4:SkySub}

The IAC Stripe 82 Legacy Project dataset has been obtained under generally photometric conditions with good seeing and low sky background. In this work, we use their sky-rectified versions of the co-added images in which the IAC team have used a non-aggressive sky subtraction strategy \citep[details on the reduction process are given in][]{TrujilloFliri2016}.

\begin{figure}
\includegraphics[width=\linewidth]{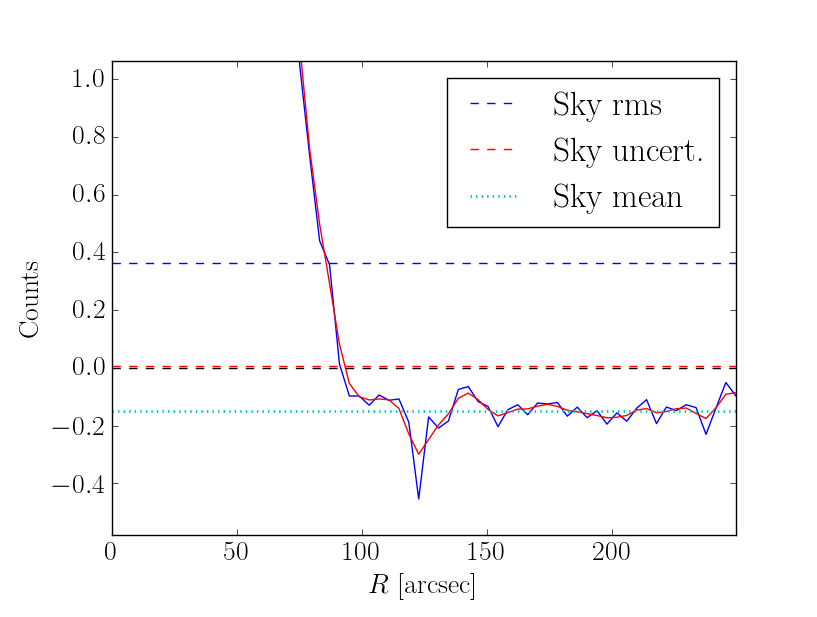} 
\includegraphics[width=\linewidth]{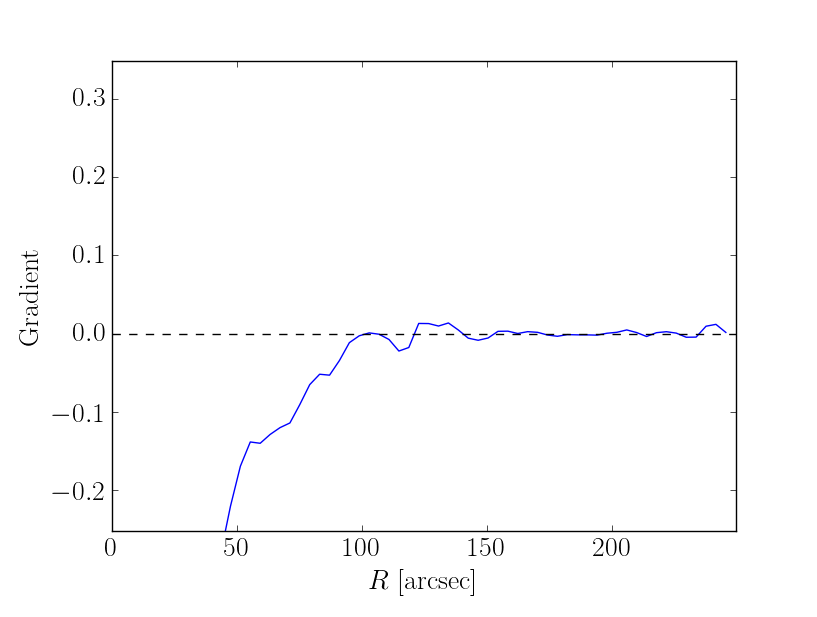}
\protect\caption[ ]{Sky subtraction for the edge-on galaxy UGC~01839 in the $r_{\rm deep}$ band of the IAC Stripe 82 Legacy Project data. Top: Outer region of the very extended radial surface brightness profile of UGC~01839 (blue solid line). Overplotted is the same profile, but convolved with a one-dimensional Gaussian kernel of 1 pixel standard deviation (red solid line), the dashed lines represent the sky rms value (blue), the uncertainty of the whole background image (red), and the zero-count position (black); the mean sky value that we subtract from the image is shown with the cyan dotted line. Bottom: Gradient function of the very extended surface brightness profile shown above; in this case, the flattest region of the gradient function was taken from 170 to 190~arcsec. UGC~01839 has an irregular background. However, there is a flat region between $\sim$170 and $\sim$190~arcsec that we use to obtain the mean sky value. Note that the sky had been oversubtracted in the original image. }
\label{figure//chap3/UGC01839:Imag_sky}  
\end{figure}

We performed a second-order (non-aggressive) background subtraction in our region of interest. To do this, we measured the background level following the procedures in \cite{PohlenTrujillo2006} and in \citet[][see details in their section~3.2]{Martinez-Lombilla2018}, and we obtained the sky value from our region of interest through very extended radial surface brightness profiles. This permits us to clearly identify where the galaxy surface brightness profile reaches the sky limit (with the current depth of our images), and consequently, where the profile starts to be dominated by the sky background component (see an example in Fig.~\ref{figure//chap3/UGC01839:Imag_sky}). It is important to note that our sample of galaxies is free of large-scale gradients in the sky background emission so that the sky level does not depend on the location of the outskirts of a galaxy \citep{Peters2017}.

\subsection{Profile extraction}\label{chap3/sec5:ProfExtr}

\begin{figure}
\includegraphics[width=.475\linewidth]{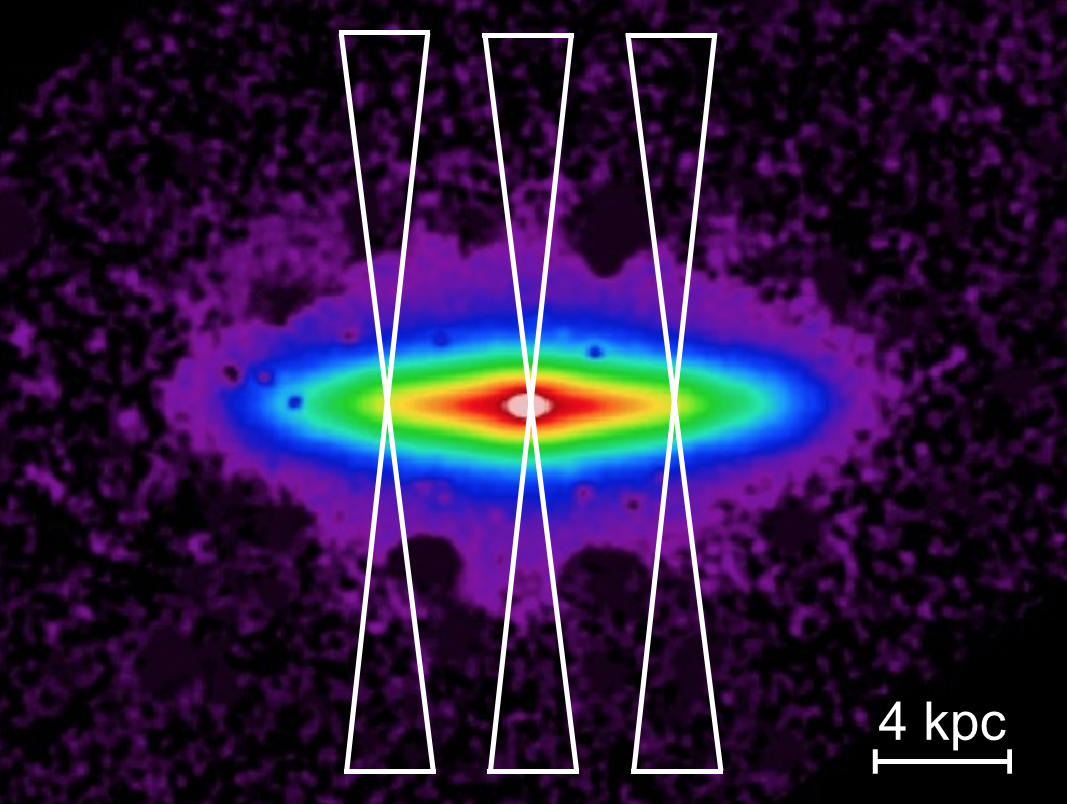}
\includegraphics[width=.525\linewidth]{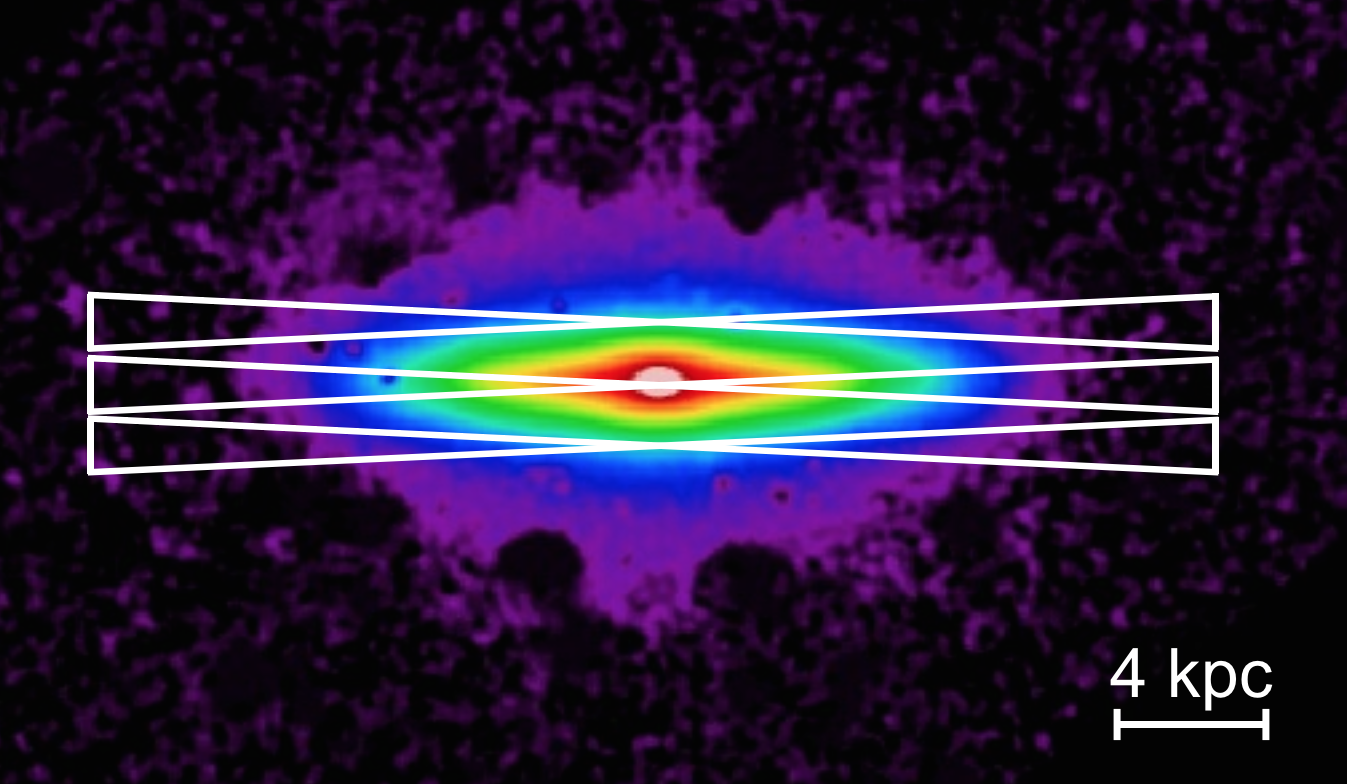}\hspace{6mm}
\protect\caption[ ]{Schematic representation of the position of the slits when the surface brightness profiles were extracted for the galaxy UGC~01040. Note that the slits were resized for each galaxy taking into account the corresponding galaxy size in order to analyse equivalent regions. Left: Vertical slits used to extract the central vertical surface brightness profiles (central slit) and the \textit{\textup{shifted}} vertical profiles (two slits on the sides). Right: Radial slits for obtaining the central radial surface brightness profile (central slit) and the \textit{\textup{shifted}} radial profiles (two slits above and below). }
\label{figure//chap3/sect3:slitsSBP} \hspace{1cm}
\end{figure}

A galaxy surface brightness profile in any given direction can be extracted by calculating the flux values through a slit along the corresponding axis (see Fig.~\ref{figure//chap3/sect3:slitsSBP} for a graphical overview of the following description). As we illustrate in Fig.~\ref{figure//chap3/sect3:slitsSBP}, the slits are double-wedge-shaped and their widths are determined in terms of the physical distance (kpc) in the galaxy, which depends on its apparent size and distance. In our study, the width of the slit increases in the direction perpendicular to the profile as we move away from the centre of the galaxy. The slit is split into logarithmic scale bins that are larger towards the outskirts. The method we followed to obtain the profiles is the same for all SDSS bands.

After the galaxy image was masked (Sect.~\ref{chap3/sect4:Masks}) and sky-subtracted (Sect.~\ref{chap3/sect4:SkySub}), we rotated it such that the plane of the galaxy disc was aligned horizontally. We began profile extraction by obtaining a central radial surface brightness profile along the galactic plane. We calculated the right- and left-side surface brightness profiles separately from the centre of the galaxy to the outer regions. We binned logarithmically in the direction of the profile and perpendicularly to it, which allowed us to increase the S/N in the outskirts of the galaxies (this causes the double-wedge shape of the slits). The final surface brightness profile was obtained by averaging the left and right radial logarithmic bins at the same distance from the centre of the galaxy. Despite small asymmetries in the radial direction, we combined the two sides to explore the surface brightness profiles down to very faint regimes (below $\mu \sim 27.0$~mag~arcsec$^{-2}$).

After producing the central radial profile, we obtained a \textit{\textup{shifted}} radial surface brightness profile at a given height above and below the galaxy mid-plane. We proceeded similarly to the central profile, but averaged the right and left radial logarithmic bins of the profile above and below the galaxy mid-plane (i.e. the average of four bins for each profile point).

The vertical surface brightness profiles were obtained in the same manner as the radial ones, but we placed a slit along the minor axis of the galaxy. We divided the vertical slit into vertical logarithmic scale bins with a variable radial width (increasing with height above or below the galaxy mid-plane), and then average the vertical logarithmic bins of the upper and the lower profiles at the same distance from the centre of the source. We extracted two vertical surface brightness profiles, one along the minor (vertical) axis of the galaxy, and a second \textit{\textup{shifted}} profile at some radial distance from the galaxy centre.

The position of the \textit{\textup{shifted}} surface brightness profiles was determined for each galaxy depending on its size. As our main scientific goal is to study the thick discs of our targets, the \textup{\textit{shifted} }radial surface brightness profile is located above or below the mid-plane but we tried to avoid (as estimated by eye) the central and inner structures such as the thin disc, the bar, the dust lane, or the bulge. At the same time, any given radial profile cannot be too far above the mid-plane, or we might start to lose S/N. We considered the S/N to be too low when the mean size of the error bars exceeded the intrinsic oscillations of the profile due to the galaxy structure (i.e. spiral arms, regions of star formation, etc). The \textit{\textup{shifted}} vertical profile follows a similar procedure to avoid the inner structures as much as possible (i.e. bulge and/or bar). The uncertainties in each bin of the profiles for all cases are defined as in \citet[][in their Section~3.3]{Martinez-Lombilla2018}.

\section{Fittings}\label{Fittings}

\subsection{Galaxy image model fitting}\label{Sect:modFitting}

As shown by \citet{Zibetti2004a}, \citet{Sandin2014}, \citet{TrujilloFliri2016}, \citet{Peters2017}, or \citet{Comeron2017}, the properties of the outskirts of a galactic disc can be strongly influenced by scattered light (see Sect.~\ref{intro}). Thus, for a given galaxy, we require a single PSF that characterises the scattered light in the region and is extended enough to cover the whole source (1.5--2 times larger than the galaxy; \citealt{Sandin2014}). In the case of the IAC Stripe 82 Legacy Project data, we already have an ultra-deep extended PSF for each field ($R \sim 8$ arcmin; see details in Sect.~\ref{chap3/sect2sub1:PSFextendedStripe 82}). The next step is to fit the galaxy in order to obtain a model without the PSF effect, and consequently, to study the true structure of the low surface brightness regions. 

 There are two principal ways of reproducing the PSF effect on the surface brightness distribution of a given object. The first approach is to directly deconvolve the original image by building a very well characterised PSF \citep[e.g.][]{Karabal2017}. The main problem of this method arises in the regions where the noise level of the image is similar to the surface brightness of the PSF, as the results in these areas can be quite uncertain. The second option consists of modelling the intrinsic light of the target and convolving it with the image PSF. The PSF-convolved model of the object is then fitted to the observed data. This process allows an observer to build a new image of the source from the combination (sum) of the PSF-deconvolved model of the target, plus the PSF-convolved fitting residuals \citep[e.g.][]{deJong2008, TrujilloFliri2016, Peters2017, Martinez-Lombilla2018}. These residuals include all the non-symmetric features, such as the spiral arms, that the model cannot fit.  
 
Our work is based on the second method listed here, as it requires fewer assumptions about the background sky and is therefore less prone to error in the low surface brightness regime. Consequently, the uncertainty in the determination of the scattered light in the outer regions of the sources is lower than in the first method. 

In order to build reliable 2D PSF-deconvolved models of our targets, we used \textsc{imfit}\footnote{Precompiled binaries, documentation, and full source code (released under the GNU Public License) are available at the following web site: \url{http://www.mpe.mpg.de/erwin/code/imfit/} .} \citep{Erwin2015}. We modelled the intrinsic light of our observed galaxy sample using a set of 2D image functions that reproduce the profiles of the galaxy components. These functions were fitted to the data by a non-linear minimisation of the total chi-squared statistic, $\chi ^{2}$, using a Levenberg-Marquardt optimisation algorithm \citep{Levenberg1944, Marquardt1963, More1978}. The surface brightness profiles of edge-on galaxies are principally divided into five components: a central bulge, a bar, a thin disc, a thick disc, and (if required) a halo. Depending on the particular case, this could be reduced to only two components (bulge + disc) or even to just a single disc. It is important to point out that \textsc{imfit} provides analytic functions that perform line-of-sight integration through 3D luminosity density models of discs seen at arbitrary inclinations. This is a very useful tool for modelling the disc structure of our edge-on galaxies. In Sect.~\ref{results} we explain in detail the particular set of functions we used for each galaxy according to their morphological classification (Table \ref{table//chap3/sect3:pysParam}). 

 After performing a reliable and comprehensive 2D fitting for each galaxy and its components, \textsc{imfit} produces the PSF-convolved and -deconvolved models. In order to quantify the effect of the PSF on the surface brightness profiles, we extracted the surface brightness profiles from the deconvolved images following exactly the same slit configurations as in the original data set. The primary benefit is that this process allows us to build a new 2D image of the observed galaxy by the combination (sum) of the PSF-deconvolved model of the target, plus the PSF-convolved fitting residuals (hereafter, deconvolved model). The deconvolved models of the galaxies are used in all subsequent steps of the analysis as the actual image of the galaxy because they are free of any background or scattered-light influence.

Although \textsc{imfit} uses a multi-threaded version of the \textsc{fftw} (fastest Fourier transform in the West\footnote{\url{http://www.fftw.org/}}) library, a relatively speedy function, adding PSF convolution to the image-fitting process does act to slow things down considerably. In addition, the high number of free parameters needed to accurately fit the functions to the observed data (around three or four, depending on the function) also makes model minimisation a time-consuming process. As a result, a large amount of time is needed for a single galaxy model fit. In the case of two or more functions, the fitting to the images was first performed for the innermost components (i.e. bulge and/or bar) and then, with the parameters for these functions fixed, the parameters of the outer components (i.e. thin and/or thick disc) were left free. The outputs for a reasonable fit from these initial runs were used to re-run the code by allowing all (or most of) the model parameters to be free. However, this issue is related to the multiple possible solutions for the modelling process. Consequently, two or more different solutions can constitute a good fit for a given galaxy. To resolve this degeneracy problem, \textsc{imfit} allows bootstrap resampling, which is a method for estimating the variability of our results by making a more detailed analysis of the parameter distributions, including potential correlations between free parameters. The combined set of bootstrapped parameter values can be used to estimate confidence intervals, and consequently, to discern between the multiple \textit{\textup{good}} solutions. However, the bootstrapping process is slow because it essentially repeats the fitting process $n$ times (although a somewhat faster version). For a reasonable and reliable set of confidence intervals at least 200 iterations are required, although 1000 or more is preferable. It is therefore better to postpone the bootstrap resampling until the final (and best) fit has been obtained, and then use it to estimate the uniqueness and the uncertainties of the fit.

 In the following sections we describe the details of an additional analytical profile-fitting method \citep{Comeron2011, Comeron2012, Comeron2014, Comeron2017} we used to obtain a more accurate estimate of the physical parameters of the galaxy discs from the deconvolved models. Our main goal is to evaluate the effect of the PSF upon the galaxy structure and to elucidate how the thick discs were formed and how they have evolved.

\subsection{Vertical luminosity profiles fitting} \label{chap3/sec6sub1:VertProfFit}

After we generated the deconvolved models of each galaxy in the sample, we fitted their vertical luminosity profiles in a similar way as described by \cite{Comeron2011, Comeron2012, Comeron2014}. In a later paper \citep{Comeron2017}, the authors explain the overall procedure in enough detail to allow reproduction or further development. We replicated their codes in Python-based routines. The main goal is to decompose a disc into the thin- and thick-disc components in order to obtain their physical parameters from the PSF-free images of the galaxies, that is, from the galaxy luminosity profiles of the deconvolved models. This procedure uses the solutions of the equations of equilibrium for two stellar and one gaseous coupled isothermal discs and therefore has the advantage of being physically motivated. There are two main reasons to make this particular vertical profile fit instead of just taking a surface brightness profile of the fitted image functions that are returned by \textsc{imfit}. The first is that the method of Comer\'on et al. of fitting vertical luminosity profiles provides an independent solution for the thin- and thick-disc decomposition and their parameters. In the case of similar solutions, it can therefore give a measure of robustness to the results. The second reason is that the physical parameters can be more accurately determined because the decomposition takes into account the gravitational interactions between the thin- and thick-disc components.

In essence, we derived the vertical height of each disc from the luminosity profiles in the vertical direction (i.e. perpendicular to the galaxy mid-planes). To do this, we fitted the vertical profiles of the data (in this case, the deconvolved model of each galaxy provided by \textsc{imfit}) by comparing them with a grid of models that are the result of a system of equations that describe three gravitationally coupled vertically isothermal discs (thin, thick, and gas discs). We thus assumed that the stellar discs are composed of two vertically isothermal discs (at a certain distance from the galaxy centre) in hydrostatic equilibrium. Both discs can interact gravitationally and have a unique vertical velocity dispersion at all heights. The third is a non-stellar (gas) disc with a given mass but no optical luminosity. The system of equations that describes the vertical density distribution for a point at a fixed height $z$ above and below the galaxy mid-plane can be written as follows \citep[adopting the formalism from][]{Narayan2002}: 

\begin{equation} \label{eq:sistEq}
\dfrac{d^{2} \rho_{i}}{d z^{2}}  = \dfrac{\rho_{i}}{\sigma^{2}_{i}} \left[ -4 \pi G (\rho_{\rm t} + \rho_{\rm T} + \rho_{\rm g}) + \dfrac{d K_{{\rm DM}}}{dz}  \right] + \dfrac{1}{\rho_{i}} \left( \dfrac{d \rho_{i}}{dz} \right)^{2}    \, \,,  
\end{equation} 

\noindent
where $\rho_{i}$ denotes the density volume of either the thin (subscript $\rm t$), thick ($ \rm T$), or gas ($\rm g$) disc; $\sigma$ is the velocity dispersion; and $K_{{\rm DM}}$ is a term related to the behaviour of the dark matter halo. However, we did not take into account the last term ($d K_{{\rm DM}}/dz = 0 $) in our fitting as it only introduces a small bias \citep[in the worst case, $M_{\rm T}/M_{\rm t}$ could be overestimated by $\sim$10\,\%][]{Comeron2011, Comeron2012}. This allowed us to assume cylindrical symmetry for the galaxy, which considerably simplifies the numerical treatment. As we are mainly interested in the study of the luminosity profiles, the gas component of the disc was not considered in this work in order to simplify the process by reducing the number of free parameters, so $\rho_{rm g} = 0$. The galaxy disc components were then fitted by considering that the thin and thick discs are two stellar fluids in hydrostatic equilibrium. However, a downside of this method is that the fitted function is not analytical and needs to be solved by numerical integration.

The luminosity profiles have a certain width given by the radial bins of $0.2 \, r_{25} < |R| < 0.5 \, r_{25}$ and $0.5 \, r_{25} < |R| < 0.8 \, r_{25}$ on each side of the bulge, where $r_{25}$ is the radius of the 25~mag~arcsec$^{-2}$ level in the $B$ band, taken from the Hyperleda database \citep[][]{Makarov2014}. To determine where to stop fitting in height above and below the galaxy mid-planes, we established a 20\,\% threshold for the maximum contribution of the error of the bin (${\rm error_{bin}}$) to its total amount of light (${\rm \# counts_{bin}}$):

\begin{equation} \label{eq:stop_fit}
\dfrac{{\rm error_{bin}}}{{\rm \# counts_{bin}}} > 0.2  \, \,. 
\end{equation}

This method for cutting the fit is different from that adopted in \cite{Comeron2011, Comeron2012, Comeron2014, Comeron2017}. We used robust means to obtain the mean value of each bin, therefore we minimised the noise in the fainter regions of the surface brightness profiles. Furthermore, we solved the problem with ``bright haloes'' because the functional shape of the disc components drops quickly and the very outer regions with excess light were not considered. It is also important to mention that the fit was made over the profile in magnitudes instead of number of counts in order to properly weight differing galaxy structures. The primary difference between our version and the code of Comer\'on et al. is that we do not need to deconvolve the observed data with the PSF of the images because we have already taken this into account (see details in Sect.~\ref{Sect:modFitting}). Finally, we note that if the scale heights of the thin and the thick discs remain roughly constant and the scale lengths of both discs are similar, the line-of-sight integration is not a problem when Eq. \ref{eq:sistEq} is applied.

\subsubsection{Mass-to-light ratio} \label{chap3/sec6sub1sub1:M/L} %(Comeron +2011)

The decomposition of galaxy discs into two gravitationally coupled vertically isothermal discs by solving the system of equations above (Eq. \ref{eq:sistEq}) is based on volume densities as a function of height. However, in our data, we measured surface brightness rather than volume density. We required the mass-to-light ratio ($\Upsilon_{i} =M_{i}/L_{i}$) of the components of our targets, that is, the thin and thick discs, to convert densities into light. In addition, structural decomposition provides the ratio of disc masses ($M_{\rm T}/M_{\rm t}$). It is therefore important to be cautious with the assumed value of mass-to-light ratios of these components, $\Upsilon_{\rm T} / \Upsilon_{\rm t}$, because this quantity contributes most to the uncertainty in the fit. 

\cite{Comeron2011, Comeron2012, Comeron2014} discussed the advantages and disadvantages of differing star formation histories (SFHs) with the purpose of obtaining the mass-to-light ratio of a given galaxy using SFHs for the thin and thick disc of the Milky Way. They used a fixed SFH with a constant $\Upsilon_{\rm T} / \Upsilon_{\rm t}$ for all the galaxies, independent of their individual properties. To avoid making this assumption, we used the relation between the photometric broad-band colour and the mass-to-light ratio from \cite{Bell2001},

\begin{equation} \label{eq:MLcolor}
\log_{10} (M/L) = a_{\lambda} + b_{\lambda} \, \times \,  {\rm colour}  \,\, ,
\end{equation}

\noindent
where $a_{\lambda}$ and $b_{\lambda}$ are coefficients taken from \cite{Roediger2015}, which provide linear least-squares fits in the SDSS bands to the mass-to-light versus colour relations (MLCRs) created on the basis of the spectral energy distributions by \cite{Bruzual2003}. Equation \ref{eq:MLcolor} allows us to obtain the mass-to-light ratio for each disc component $M_{i}/L_{i}$ by taking radial surface brightness profiles along the thin and the thick disc separately and obtaining the corresponding colour profiles.

\subsubsection{Physical parameters of thin and thick discs} \label{chap3/sec6sub1sub2:PhysParam} 

From the vertical fitting (Sect.~\ref{chap3/sec6sub1:VertProfFit}) of the luminosity profiles of the deconvolved model of each galaxy, we estimated a set of physical parameters that allowed us to analyse the thin- and thick-disc properties. These quantities help to discern between the various disc formation scenarios and are described as follows:

\begin{itemize}
\item Scale heights: the scale height of the thin- and thick-disc components ($z_{\rm 0t}$ and $z_{\rm 0T}$, respectively) is defined as the vertical distance (in arcsec) over which the density of a particular physical variable (the luminosity obtained from the profile in this case) drops by a factor of $e$ ($\sim$2.718) from its centre and it is heavily dependent on the type of object. For each bin the scale height was determined by establishing the distance between the location where the flux is a factor 100 lower than in the galaxy mid-plane and the position where the flux is 100$\times e$ lower than at the mid-plane in the fitted surface brightness profiles for each of the two stellar disc components.

\item Mid-plane mass-density ratio: the ratio of the thick-disc over the thin-disc mass density in their central regions ($\rho _{\rm 0T}/\rho _{\rm 0t}$). This parameter is directly obtained from the fit, as the solution of Eq.~\ref{eq:sistEq} yields the corresponding values of $\rho_{\rm 0T}$ and $\rho _{\rm 0t}$ for each galaxy.

\item Thick- to thin-disc mass ratio: the ratio of the cumulative mass density (column mass) along the profile of each disc component, that is, a surface-density ratio ($\Sigma \rho_{\rm T}/ \Sigma \rho_{\rm t} = \Sigma_{\rm T}/ \Sigma_{\rm t}$).

\end{itemize}

\section{Results}\label{results}

In this section, we present the main results of our fits for each galaxy individually in the $r_{\rm deep}$ band of the IAC Stripe 82 Legacy Project (i.e. the SDSS \textit{g, r,} and \textit{i} combined band). The aim is to compare the observed data and the corresponding PSF-deconvolved model to evaluate the PSF influence using surface brightness profiles in both the radial and vertical directions (as explained in Sect.~\ref{chap3/sec5:ProfExtr}). It is crucial to perform the PSF-deconvolution in order to work with what is assumed to be the actual flux from these galaxies. We therefore analysed the radial and vertical surface brightness profiles for the observed data, the PSF-convolved model, and the PSF-deconvolved model. For each model, the corresponding galaxy components were also studied. We performed a bootstrap resampling (described in Appendix~\ref{appendix}) for each 2D galaxy model using the astronomical software \textsc{imfit} \citep{Erwin2015}. The main reason for a bootstrap resampling is that there is a strong degeneracy for the disc parameters with one another, so that multiple possible solutions may appear for a given galaxy in the modelling process.

After we obtained the PSF-deconvolved image (plus the residuals of the PSF-deconvolved fit), we fitted its vertical light distribution by solving a system of equations that describe two gravitationally coupled vertically isothermal discs (thin and thick discs), as described in \citet[][]{Comeron2011, Comeron2012, Comeron2014, Comeron2017}, and in Sect.~ \ref{Sect:modFitting}. The main goal was to decompose the disc of the PSF-deconvolved model into the thin- and thick-disc components considering their interactions, in order to obtain their main physical parameters such as the scale heights ($z_{{\rm t}}$ and $z_{{\rm T}}$, respectively), the initial mass density ratio ($\rho _{\rm 0T}/\rho _{\rm 0t}$), or the thick- to thin-disc mass ratio ($\Sigma_{{\rm T}}/\Sigma_{{\rm t}}$). The weighted mean values of all these physical parameters of the thin and thick discs of each galaxy are shown in Table~\ref{table//chap3/sect3:pysParam}.
All fitted surface brightness profiles for the whole galaxy sample are shown in \citet[][]{LombillaPhD}.

\subsection{Intermediate- to high-mass galaxies: NGC~0429 and UGC~01040} \label{chap3/intermMass}

\begin{figure*}
\begin{center}
\includegraphics[width=150mm]{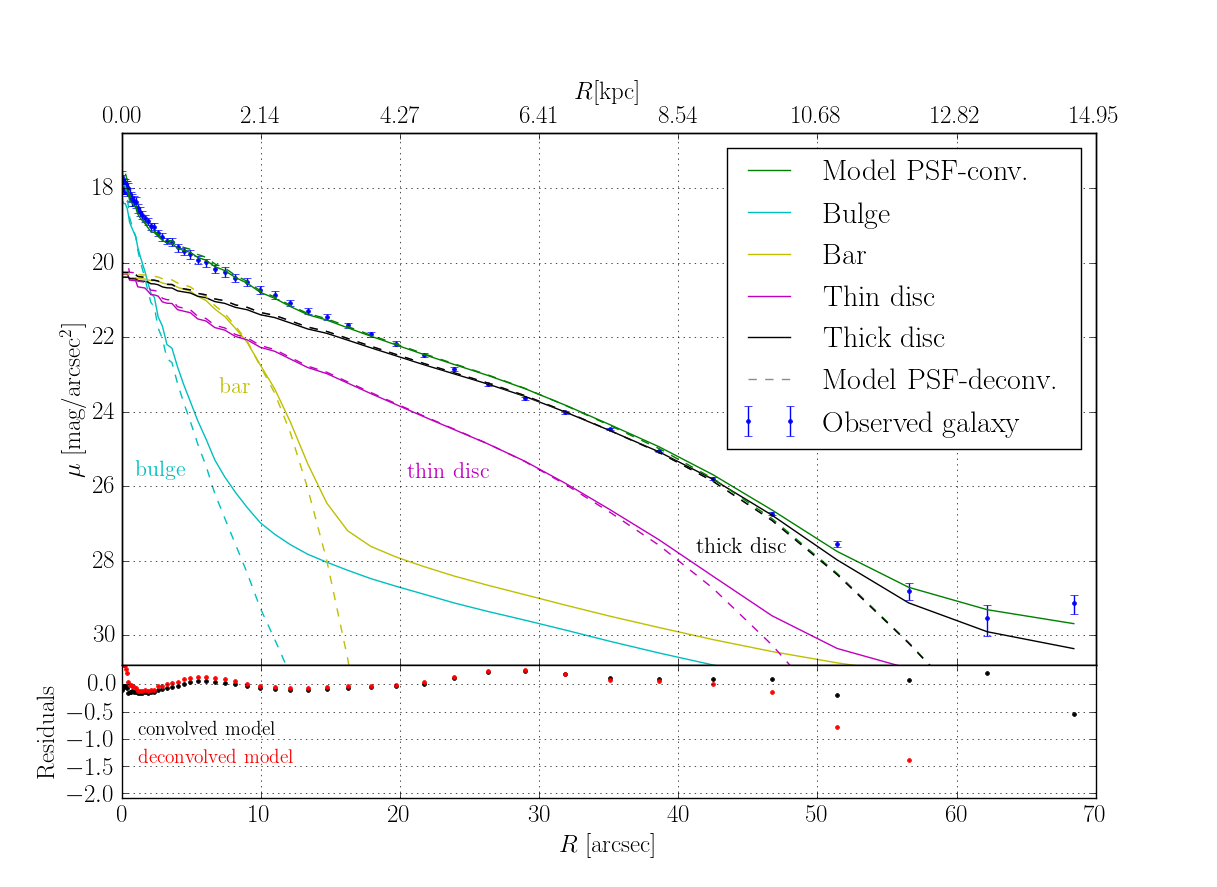}
\includegraphics[width=90mm]{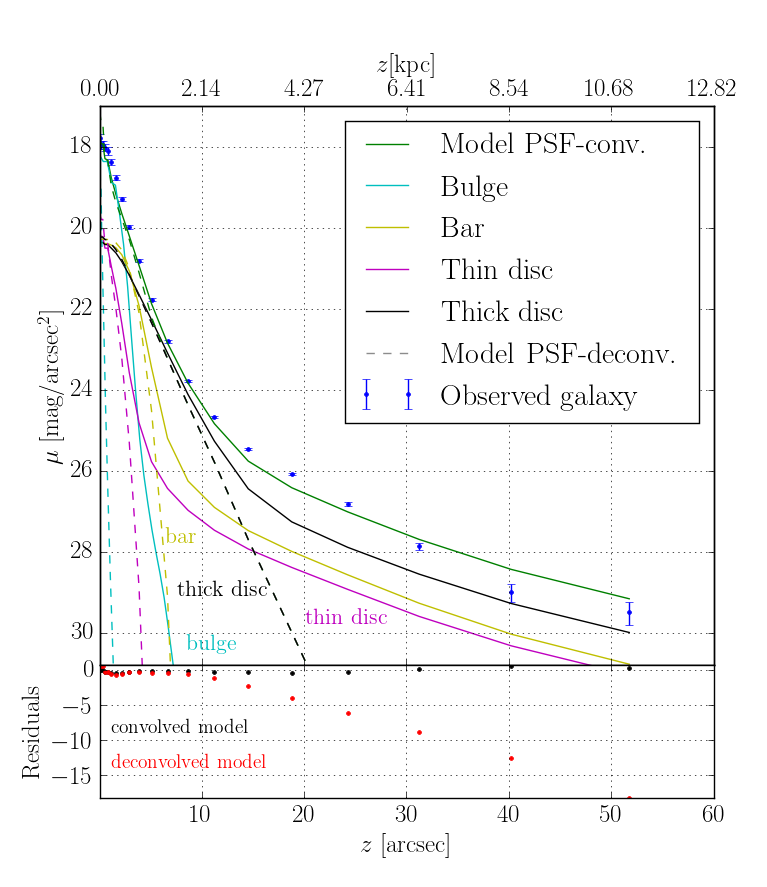} 
\protect\caption[ ]{Top: Radial surface brightness profiles along the mid-plane of the edge-on galaxy UGC~01040 in the combined $r_{\rm deep}$ band of the IAC Stripe 82 Legacy Project data. We plot our data with blue dots, the PSF-convolved model and its components are shown with solid curves, and the PSF-deconvolved model and its corresponding components are the dashed curves. The whole PSF-convolved (and -deconvolved) model is represented in green, and the components follow the colour code of the legend. In the lower panel, we show the difference between the observed data and either the PSF-convolved model (black points) or the PSF-deconvolved models (red points). Bottom: Same as in the top panel, but now we show the vertical surface brightness profiles along the minor axis for the edge-on galaxy UGC~01040. The whole set of plots, including the shifted radial and vertical profiles for all the galaxy sample, have been published in \citet[][]{LombillaPhD}.}
\label{figure//chap3/UGC01040:RadVertSBP}   
\end{center}
\end{figure*}

%\begin{figure}
%\begin{center}
%\includegraphics[width=\linewidth]{minor_profile_KpcArcsec_grideep_highMass} 
%\protect\caption[ ]{As Fig.~\ref{figure//chap3/UGC01040:RadSBP}, but now we show the vertical surface brightness profiles along the minor axis for the edge-on galaxy UGC~01040. The whole set of plots, including the shifted vertical profiles for all the galaxy sample have been published in Chapter~3 of \citet[][]{LombillaPhD}}
%\label{figure//chap3/UGC01040:VertSBP}
%\end{center}
%\end{figure}

NGC~0429 and UGC~01040 (see optical colour images from the SDSS DR12 in Fig.~\ref{figure//chap3/Ind_galaxies}) are lenticular galaxies of intermediate to high stellar mass (4.4-8.3$\times$10$^{10}$~$M_{\odot}$ and 1.3-2.8$\times$10$^{10}$~$M_{\odot}$, respectively). NGC~0429 is the most distant galaxy in our sample at more than twice the distance to UGC~01040, the next galaxy\ (see Table~\ref{table//chap3/sect3:finalSample}). Table~\ref{table//chap3/sect3:finalSample} also shows that it is unclear whether UGC~01040 has a bar, but our models return very good fits when a bar is included.

In Fig.~\ref{figure//chap3/UGC01040:RadVertSBP} we show an example of the radial and vertical surface brightness profiles for the observed data of UGC~01040, the PSF-convolved (and -deconvolved) model, and their components. In this particular case, it was necessary to use four galaxy components to find a reliable fit: a bulge using a 2D S{\'e}rsic function, a bar represented by an elliptical 2D S{\'e}rsic function using generalised ellipses (``box'' to ``disc'' shapes) for the isophotes instead of pure ellipses \citep{Athanassoula1990}, and finally, a thin and a thick disc using a 3D broken exponential disc function. For NGC~ 0429, we required three components: a bulge based on a 2D S{\'e}rsic function, and both a thin and a thick disc, but this time using a 3D pure exponential disc function \citep[for more details about the analytical functions, see][]{Erwin2015}. This type of 2D model fit was performed for all the galaxies in the sample. 

In the case of NGC~0429, the PSF effect in the radial surface brightness profiles becomes relevant below $\sim$28~mag~arcsec$^{-2}$. Thus, if we do not account for the PSF along the radial profile, we overestimate the outer part of the galaxy but by a similar amount in all the components. For the vertical profiles, this effect is much more dramatic as it starts to be significant beyond $\sim$25~mag~arcsec$^{-2}$. Consequently, the galaxy flux and thus its mass (for a given $M/L$) is overestimated by a factor of $\sim$10 in the outskirts ($z \sim$16~kpc) because of the way in which the PSF affects the bulge and the thin disc, rather than the thick disc. Consequently, we infer that the PSF-deconvolved components of the galaxy model contribute less to the total flux when the slope of the profile is steeper (although this also depends on the central intensity).

For UGC~01040, the most complex galaxy of the sample with four structural components for the 2D model fit, the profiles do not show any remarkable discrepancy between the observed data and the PSF-convolved model (see the residual plots in the lower part of each panel in Fig.~\ref{figure//chap3/UGC01040:RadVertSBP}), except for a slight excess of light in the data in comparison with the PSF-convolved model at $z \sim$2-4~kpc in the shifted vertical surface brightness profile, probably due to some asymmetry. The PSF effect in the radial direction (top panel Fig.~\ref{figure//chap3/UGC01040:RadVertSBP}) becomes relevant below $\sim$28~mag~arcsec$^{-2}$, producing a slight overestimate in the outer part of the galaxy if not accounted for. Once again, the PSF has a dramatic effect on the vertical light distribution in the galaxy beyond $\sim$25~mag~arcsec$^{-2}$ (bottom panel Fig.~\ref{figure//chap3/UGC01040:RadVertSBP}). The galaxy UGC~01040 is most affected by the PSF in the sample. The main reason is the very steep slope of the vertical surface brightness profiles. However, in this case, the dominant component in the outer regions of the PSF-convolved model is the thick disc, although the thin disc and the bar also contribute significantly, mainly in the case of the minor axis profile. After the PSF is removed, the vertical light distribution reaches a level of $\sim$40\,\% of the observed data, which corresponds to a decrease in flux (i.e. mass for a given $M/L$ value) of a factor of at least 40 at $z \sim$4~kpc. Thus, it is critical to account for the PSF effect for these intermediate- to high-mass galaxies as the scattered light produces an overestimation in all galaxy components.

\begin{figure*}
  \begin{center}
      \includegraphics[width=70mm, clip=true, trim = 0mm 0mm 0mm 0mm]{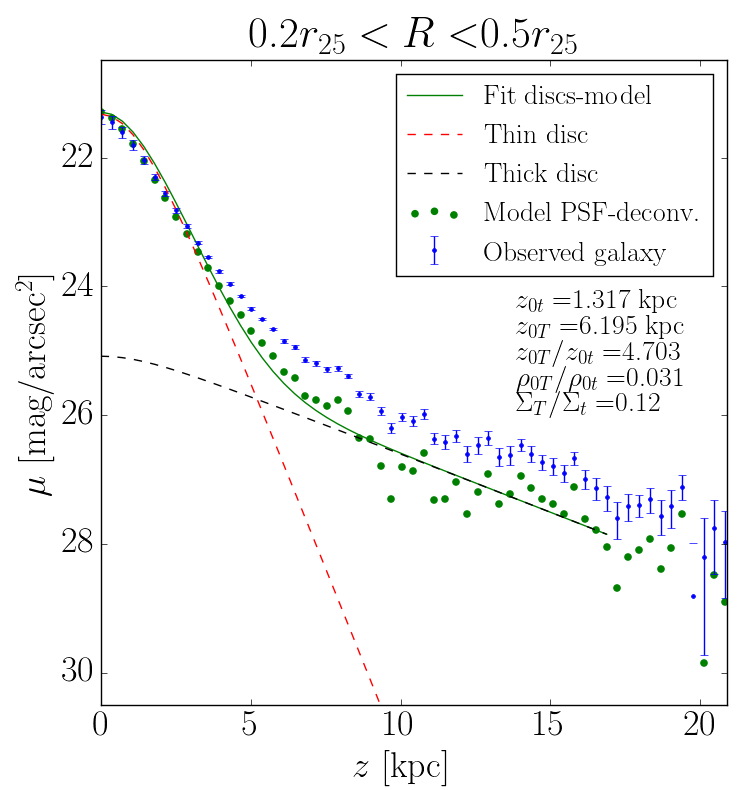}
  \includegraphics[width=70mm, clip=true, trim = 9mm 202mm 185mm 7mm]{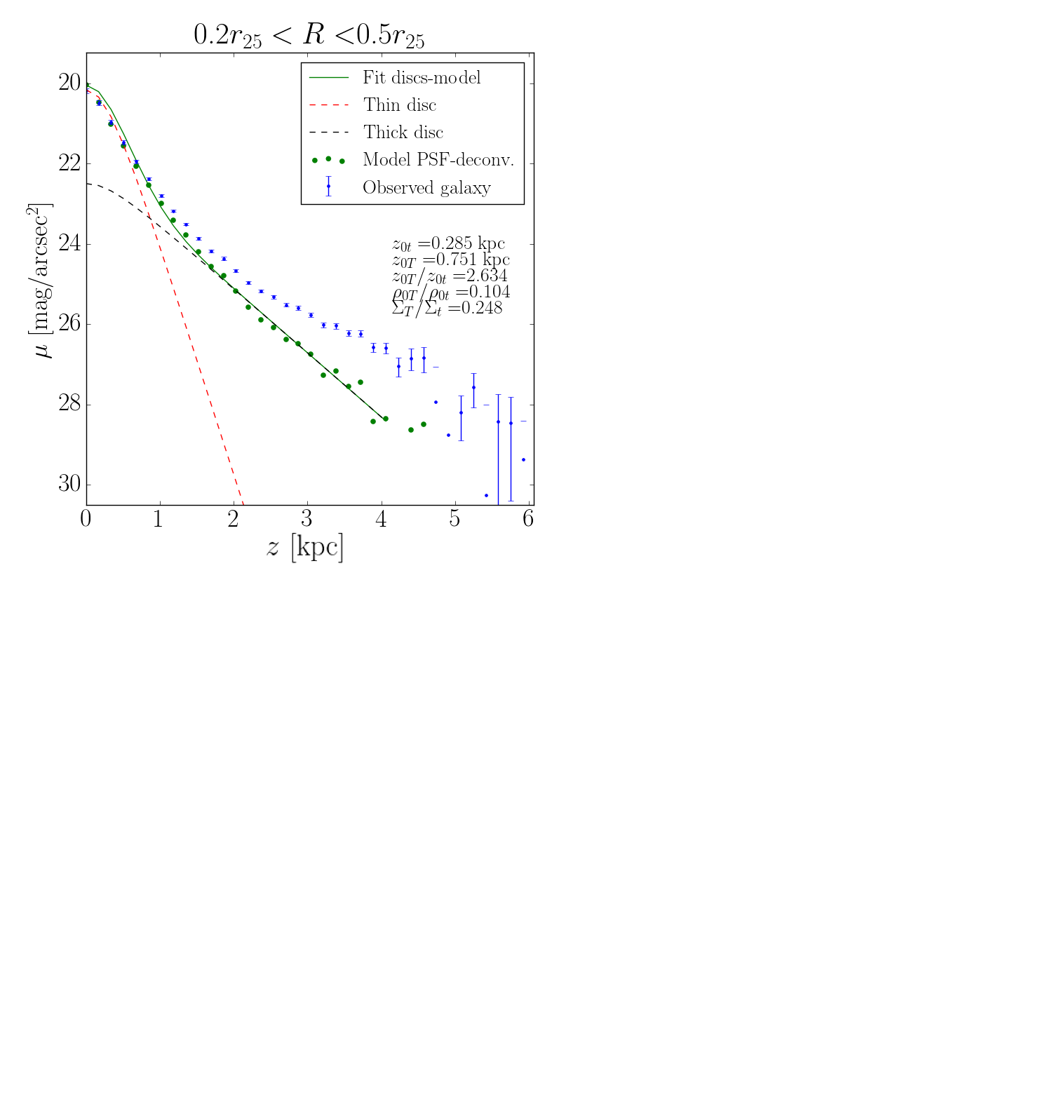}
      \includegraphics[width=70mm, clip=true, trim = 5mm 0mm 0mm 0mm]{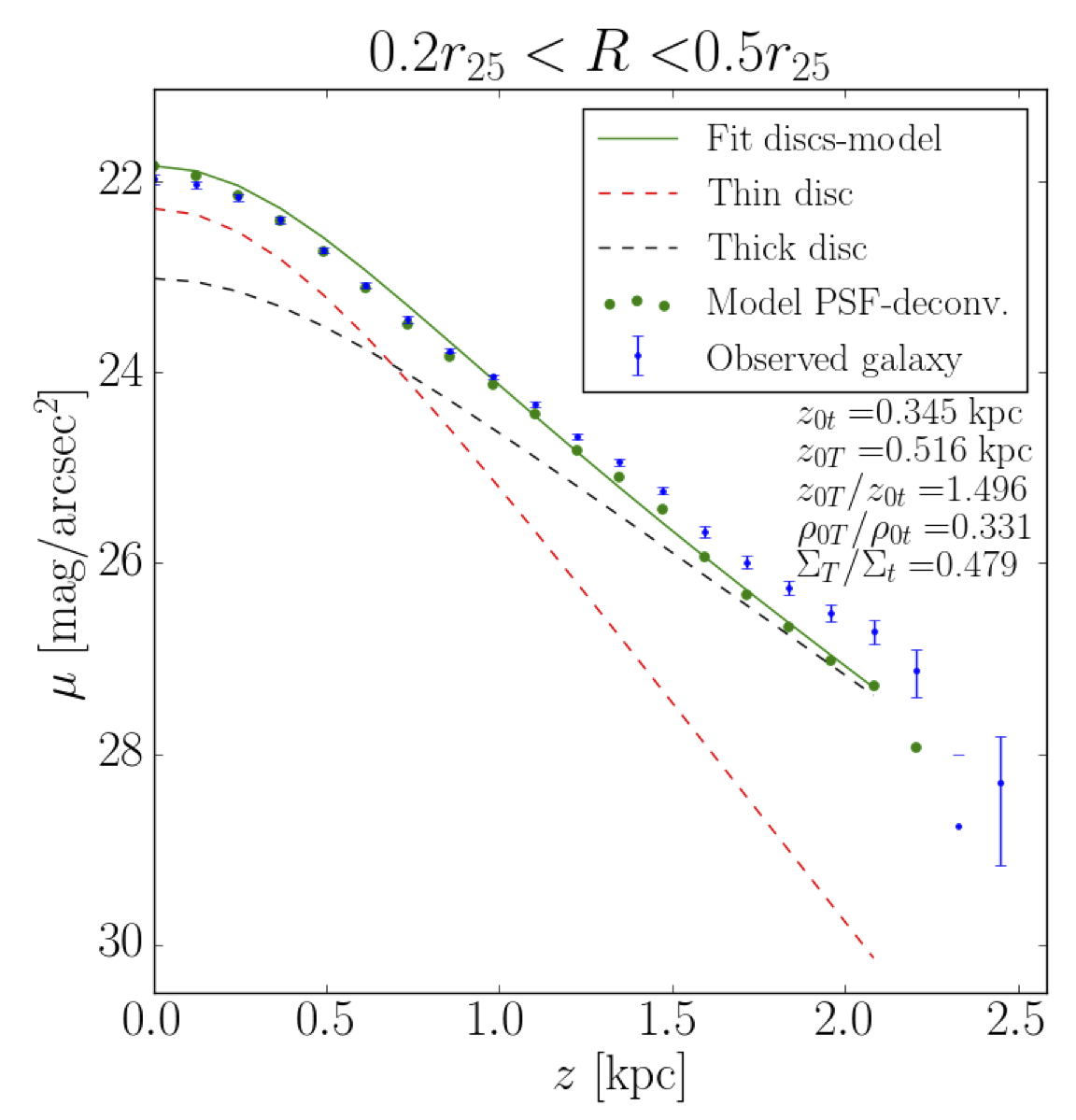}
  \includegraphics[width=70mm, clip=true, trim = 9mm 202mm 185mm 7mm]{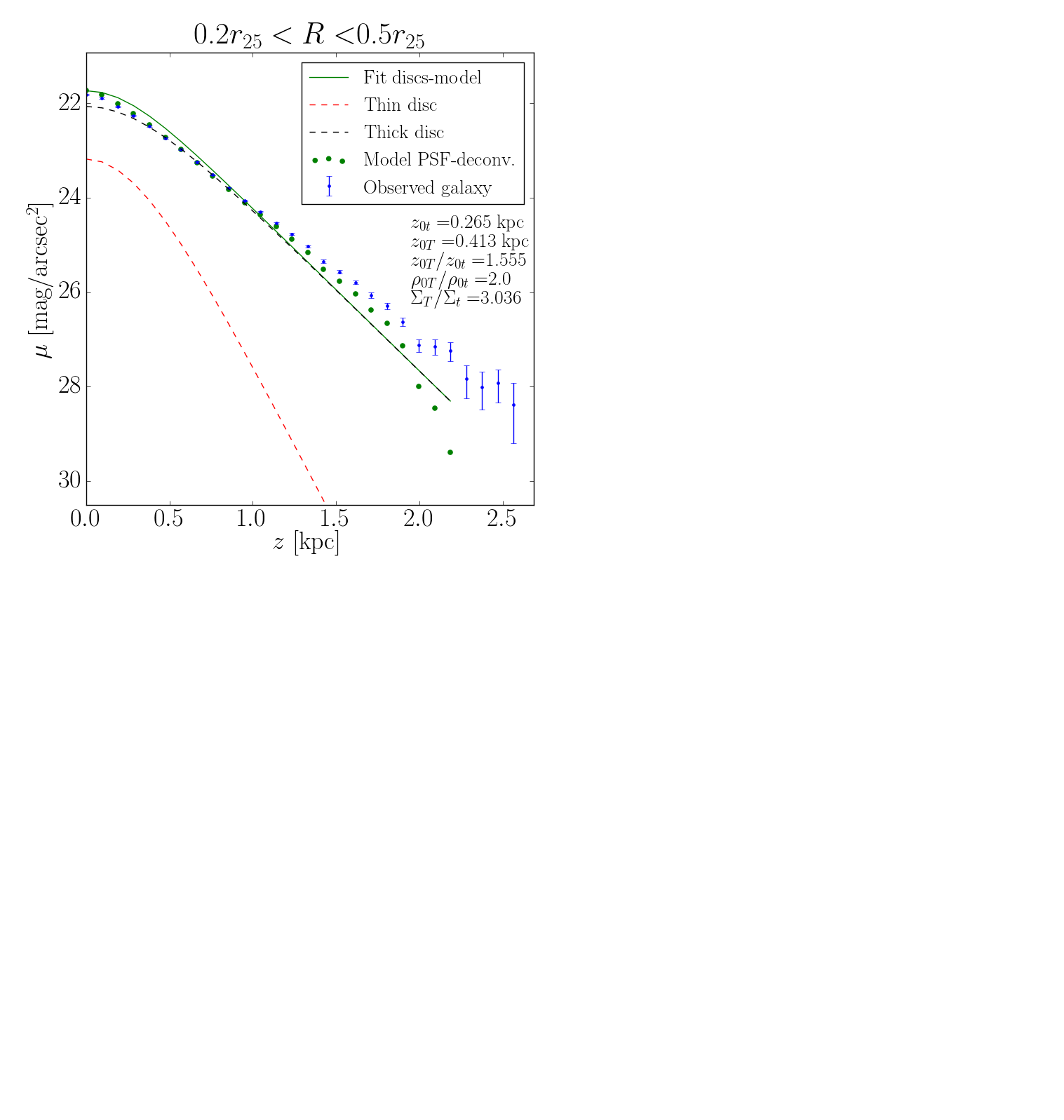}
    \includegraphics[width=70mm, clip=true, trim = 9mm 202mm 185mm 7mm]{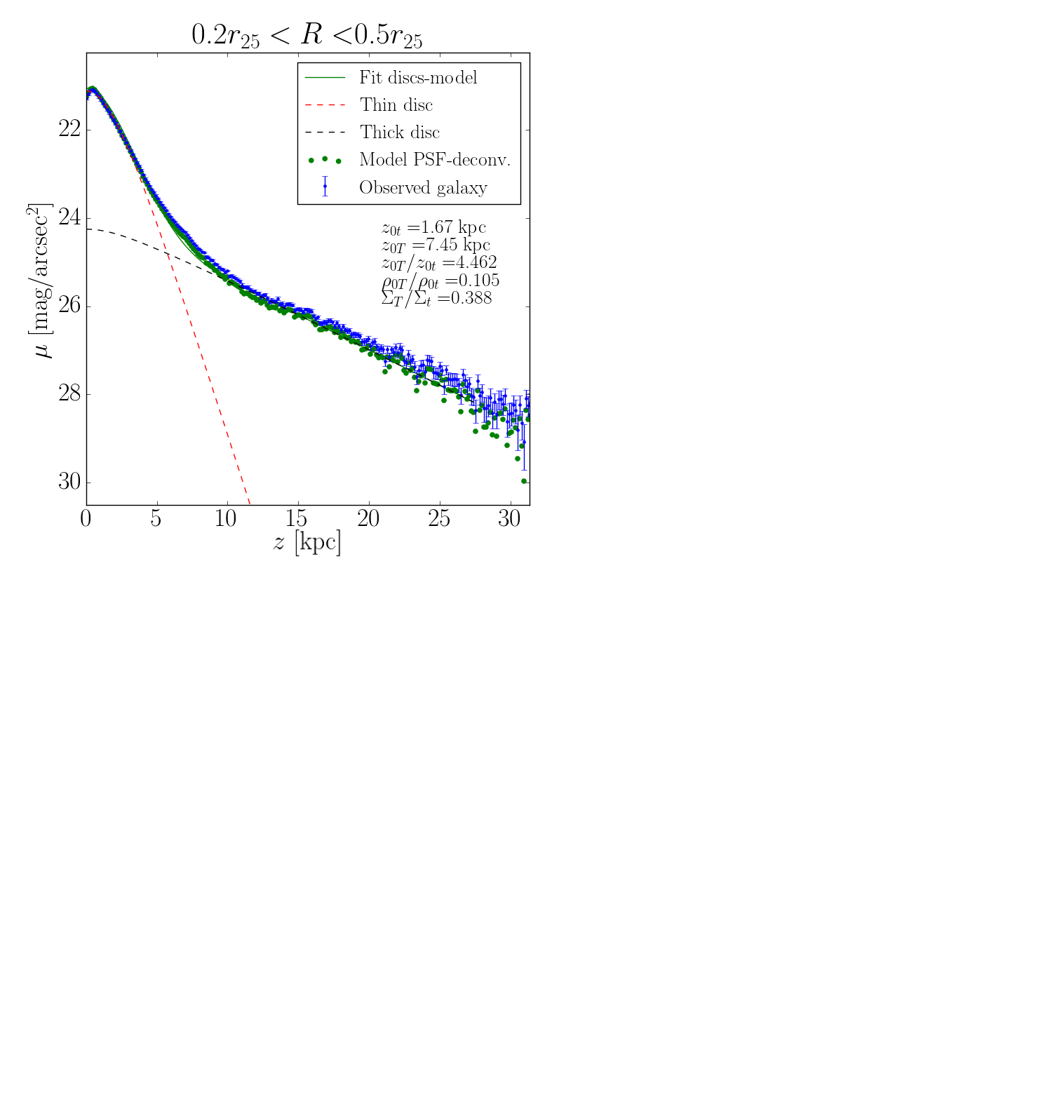}
  \caption{Vertical surface brightness profiles fits to the PSF-deconvolved models in one radial bin for the whole edge-on galaxy sample in the combined $r_{\rm deep}$ band. From top to bottom and from left to right: the two intermediate- to high-mass galaxies (NGC~0429 and UGC~01040), the two low-mass galaxies (UGC~00931 and UGC~01839), and NGC~1032. The full procedure for obtaining the fits is explained in Sect.~\ref{chap3/sec6sub1:VertProfFit}. The observed data are the blue points, the PSF-deconvolved models are the green filled circles, and the resulting fits are the green solid lines. The dashed lines indicate the thin- (red) and thick-disc (black) contributions to the observed data. We show the physical parameters obtained from the fits of the bin for each galaxy: the scale heights of the thin and the thick discs ($z_{{\rm t}}$ and $z_{{\rm T}}$, respectively); the fitted initial mass density ratio ($\rho _{\rm 0T}/\rho _{\rm 0t}$); and the thick- to thin-disc mass ratio ($\Sigma_{{\rm T}}/\Sigma_{{\rm t}}$) in the bin. The PSF contribution over the galaxy NGC~1032 is almost negligible due to the smooth light distribution. The whole set of plots, including the four radial bins, has been published in \citet[][]{LombillaPhD}.}
  \label{figure//chap3/ComeronDeconv}
  \end{center}
\end{figure*}

We proceed to analyse the results of the vertical fits to the light distribution through the resolution of a system of equations that describe gravitationally coupled vertically isothermal discs (thin and thick discs), as described in \citet[][and Sect.~\ref{chap3/sec6sub1:VertProfFit} of this work]{Comeron2011, Comeron2012, Comeron2014, Comeron2017}. Figure~\ref{figure//chap3/ComeronDeconv} shows the thin- and thick-disc decomposition for the fit of the PSF-deconvolved model at a given radial distance of UGC~01040. In comparison with previous 2D models, these fits just reach surface brightness values above $\sim$27-27.5~mag~arcsec$^{-2}$, depending on the bin, and seem to have problems when the profile is more noisy or when it presents asymmetries. These fits were not considered in our results.

We note that after we added the residuals of the PSF-convolved model to the PSF-deconvolved galaxy image (i.e. green point profiles in Fig.~\ref{figure//chap3/ComeronDeconv}), the dramatic effect of the PSF diminished along the vertical surface brightness distribution in comparison with the bottom panel of Fig. \ref{figure//chap3/UGC01040:RadVertSBP}, but it still adds artificial light to the galaxy outskirts. The difference between the observed data and the PSF-deconvolved model is now $\sim$1~mag~arcsec$^{-2}$ at $z \gtrsim$10~kpc for NGC~0429, and $\sim$1.5~mag~arcsec$^{-2}$ at $z \sim$4~kpc for UGC~01040, which corresponds to an excess of galaxy flux, and therefore mass, of a factor of $\sim$2.5 and $\sim$4, respectively. This is a critical result that reveals just how crucial careful PSF characterisation is.

\begin{table*}
\begin{center}
\caption{Physical parameters of the thin and thick discs of each galaxy in our sample.}\vspace{2mm}
\label{table//chap3/sect3:pysParam} 
\begin{tabular}{c|cc|cc|cc|cc|cc}
Phys. Param. & \multicolumn{2}{c}{NGC}{0429 } & \multicolumn{2}{|c}{NGC}{01040} & \multicolumn{2}{|c}{UGC}{00931} & \multicolumn{2}{|c}{UGC}{01839} & \multicolumn{2}{|c}{UGC}{1032} \\
\hline
$z_{\rm t}$ [kpc] & 1.14 & 1.38 & 0.31 & 0.36 & 0.31 & 0.35 & 0.31 & 0.28 & 1.74 & 1.8 \\
$z_{\rm T}$ [kpc] & 5.15 & 5.75 & 1.13 & 0.91 & 0.61 & 0.53 & 0.6 & 0.47 & 8.82 & 8.81 \\
$z_{\rm T}/z_{\rm t}$ & 4.51 & 4.17 & 3.64 & 2.53 & 1.96 & 1.51 & 1.93 & 1.68 & 5.07 & 4.89 \\
$\rho_{\rm 0T}/\rho_{\rm 0t}$& 0.09 & 0.04 & 0.16 & 0.09 & 0.24 & 0.33 & 0.29 & 1.31 & 0.11 & 0.13 \\
$\Sigma_{\rm T}/\Sigma_{\rm t}$ & 0.34 & 0.15 & 0.5 & 0.22 & 0.45 & 0.46 & 0.51 & 2.19 & 0.42 & 0.49 \\
$ \Upsilon_{\rm T} / \Upsilon_{\rm t}$ & - & 0.99 & - & 1.07 & - & 1.97 & - & 1.46 & - & 0.56 \\
\end{tabular}

\end{center}
\tablefoot{Weighted mean values of the physical parameters of thin and thick discs of each galaxy for the observed data (left values of each column) and for the PSF-deconvolved models (right values of each column). These results are obtained from the vertical fits of the light distribution determined by solving a system of equations that describe two gravitationally coupled vertically isothermal discs (thin and thick discs). The mass-to-light ratios between the thin and thick discs ($(M/L)_{\rm T}/(M/L)_{\rm t} = \Upsilon_{\rm T} / \Upsilon_{\rm t}$) is only obtained for the PSF-deconvolved models. For more details about the parameters, see Sect.~\ref{chap3/sec6sub1sub2:PhysParam}.}
\end{table*}

For each galaxy we obtained a set of physical parameters that are shown in Table~\ref{table//chap3/sect3:pysParam}. We did not derive uncertainties because they are very difficult to obtain due to the complicated fitting process (we solve a system of equations that describe two gravitationally coupled discs). The reliability of the fits is well studied in \cite{Comeron2011}, therefore we use the same number of significant digits. In our case, the deviations between the different values of the physical parameters in the panels of each of those figures may be representative of the uncertainties in the method.

It is important to note that in the case of NGC~0429, against the expectations, the vertical scale height of the thick disc ($z_{{\rm T}}$) is larger after the PSF is accounted for, so that the thick disc becomes thicker. The ratio between the observed data and the PSF-deconvolved thick- to thin-disc mass ratios (i.e. $(\Sigma_{{\rm T}}/\Sigma_{{\rm t}})_{\rm obs} / (\Sigma_{{\rm T}}/\Sigma_{{\rm t}})_{\rm PSFdec}$) is equal to 2.3, which is in relatively good agreement with the previously estimated excess of mass (a factor of $\sim$2.5). In UGC~01040, the vertical scale of the thick disc ($z_{{\rm T}}$) is lower when the PSF is accounted for, so the vertical distribution of the thick disc becomes steeper, in full agreement with expectations. The ratio between the observed data and the PSF-deconvolved thick- to thin-disc mass ratios (i.e. $(\Sigma_{{\rm T}} / \Sigma_{{\rm t}})_{\rm obs} / (\Sigma_{{\rm T}}/\Sigma_{{\rm t}})_{\rm PSFdec}$) is equal to 2.3, which is lower than the previously estimated excess of mass (a factor of $\sim$4). In any case, both values are high enough to claim that a proper PSF treatment of the observed data is absolutely indispensable.
The thin- and thick-disc components may have different shapes in the vertical luminosity profile fits (e.g. Fig.~\ref{figure//chap3/ComeronDeconv}) than in the 2D models (e.g. bottom panel in Fig.~\ref{figure//chap3/UGC01040:RadVertSBP}) because in the first physical interactions between the thin and thick discs are considered.

\subsection{Low-mass galaxies: UGC~00931 and UGC~01839} \label{chap3/lowMass}

UGC~00931 (optical colour image from the SDSS DR12 in Fig.~\ref{figure//chap3/Ind_galaxies}) is the galaxy with the lowest inclination in the sample and also the least massive (3.1$\times$10$^{9}$~$M_{\odot}$), while UGC~01839 (colour image in Fig.~\ref{figure//chap3/Ind_galaxies}) is a low-mass (5.9$\times$10$^{9}$~$M_{\odot}$) late-type spiral and the closest galaxy in the sample (see details in Table~\ref{table//chap3/sect3:finalSample}). Visually, the disc of UGC~00931 is very diffuse and asymmetrical, which leads to enhanced residuals in the 2D model. UGC~01839 also has an asymmetric morphology with diffuse and broken arms, and no central bulge. However, as indicated in Table~\ref{table//chap3/sect3:finalSample} with the letters ``sp'' after the morphological type, its classification has considerable uncertainty because its internal structures are not easily distinguished due to the high inclination of the galaxy.

\begin{figure}
\begin{center}
\includegraphics[width=\linewidth]{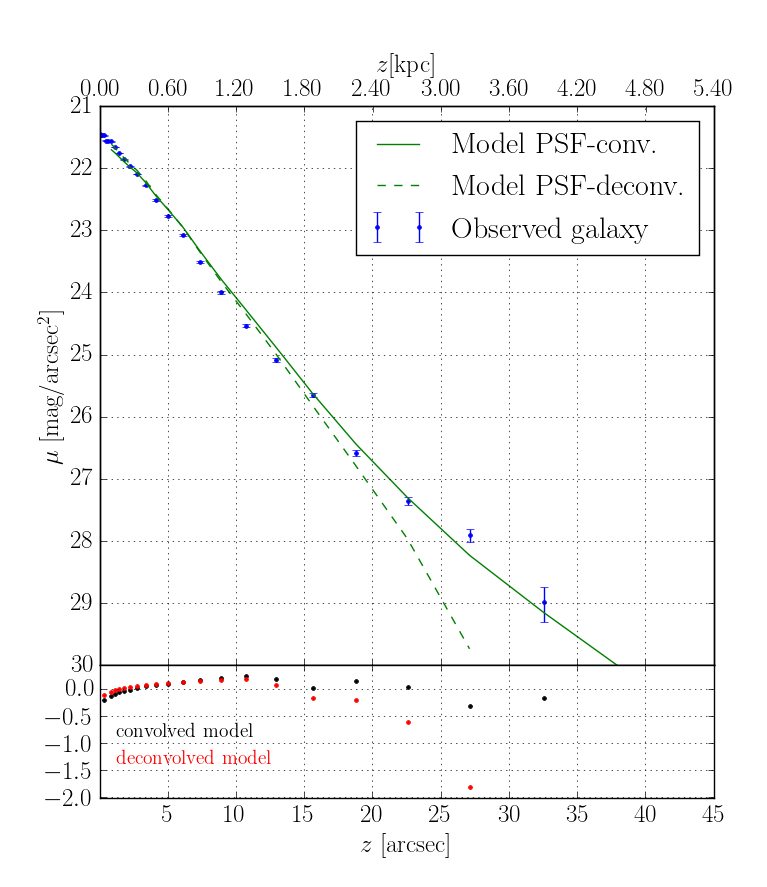} 
\protect\caption[ ]{Vertical surface brightness profiles as in the lower panel of Fig.~\ref{figure//chap3/UGC01040:RadVertSBP}, but for the edge-on galaxy UGC~01839. In this case, just one disc component was needed to obtain a reliable model. }
\label{figure//chap3/UGC01839:VertSBP}
\end{center}
\end{figure}

We only required a single 3D broken exponential disc for a satisfactory fit of these two low-mass edge-on galaxies. Thus, no thick (or thin) disc component was necessary to model them. To illustrate this, the vertical surface brightness profile of UGC~01839 and its 2D model are shown in Fig.~\ref{figure//chap3/UGC01839:VertSBP}. This vertical surface brightness profile is affected by the PSF when values below $\sim$26~mag~arcsec$^{-2}$ ($z \sim$2~kpc) are reached. At $z \sim$3~kpc, the mass is overestimated by a factor of $\sim$3. UGC~00931 is affected in an equivalent way.

The following results are based on fits along the vertical light distribution, considering the physical interactions between two gravitationally coupled vertically isothermal discs \citep[details in][and in Sect. \ref{chap3/sec6sub1:VertProfFit} of this work]{Comeron2011, Comeron2012, Comeron2014, Comeron2017}. In Fig.~\ref{figure//chap3/ComeronDeconv} we illustrate the thin- and thick-disc decomposition for the PSF-deconvolved model of UGC~01839, down to a surface brightness of $\sim$27.5-28~mag~arcsec$^{-2}$.

By comparing these results with the previously presented 2D models (e.g. Fig.~\ref{figure//chap3/UGC01839:VertSBP}), a discrepancy becomes clear: in this case, the fit can be made using two discs instead of just one. \cite{Comeron2017} found only 17 galaxies out of 141 that showed no clear traces of a thick disc. These 17 galaxies are on average less massive than those that show at least two discs. These fits based on gravitational interactions are limited to a two-disc decomposition, so that we always obtained two components. Thus, ``no clear traces of a thick disc'' means that the result is similar to ours in the performed fits for these two low-mass galaxies, in the sense that either one disc clearly dominates the other (e.g. Fig.~\ref{figure//chap3/ComeronDeconv}), or the two discs are very similar and therefore confirm a single disc-like profile. The vertical scale height ratios of the thin and thick fitted discs are quite similar in both cases (in comparison with other galaxies in the sample, less than a factor of 2). In UGC~00931, the disc mass excess in the galaxy outskirts due to the PSF effect (taking into account all the non-symmetric features such as the spiral arms, which the model cannot fit, and the physical interactions between both discs) is now a factor of $\sim$1.5, while for UGC~01839, the disc mass is  overestimated a factor of $\sim$2. Thus, as for the two intermediate- to high-mass galaxies, the low surface brightness structures of UGC~00931 and UGC~01839 are significantly overestimated when the PSF is not considered because of the light that is scattered from the inner to the outer regions.

The weighted mean values of the physical parameters obtained for our two low-mass galaxies are shown in Table \ref{table//chap3/sect3:pysParam}. The scale heights of both the thin and thick disc ($z_{{\rm T}}$ and $z_{{\rm t}}$) are smaller when the PSF is accounted for, as expected. This is because after removing the PSF effect, the fit uses a dominant component (the thick disc) that fits the PSF-deconvolved model points, therefore its scale height should be smaller, plus a faint thin disc. These results are consistent with an increase in the ratio of thick- to thin-disc masses.

\subsection{A particular case: NGC~1032} \label{chap3/partCaseNGC1032}

NGC~1032 (see an optical colour image from the SDSS DR12 in Fig.~\ref{figure//chap3/Ind_galaxies}) is a very interesting galaxy because it is the only one in common with the S$^{4}$G-based sample of \citet{Comeron2011, Comeron2012, Comeron2014, Comeron2017}. It has a high dynamical mass (3.3$\times$10$^{11}$~$M_{\odot}$, see Table~\ref{table//chap3/sect3:finalSample}), which is isolated \citep{Tully2015}, and it has a straight and thin dust lane along its mid-plane, out to $\sim$7.5~kpc.

The radial and vertical surface brightness profiles of NGC~1032, as for NGC~0429, required three components: a bulge based on a 2D S{\'e}rsic function, and both a thin and a thick disc, but this time using a 3D pure exponential disc function \citep[see][]{Erwin2015}. The most interesting aspect in this galaxy is that the PSF does not play an important role, not even in the outermost parts of the galaxy. This is mainly because the slope of the profiles is smoother and less steep than in the other galaxies of the sample. By analysing the structural components individually, it is clear that the PSF effect is stronger in the bulge and thin-disc components than in the thick disc, and along the vertical direction, although the thick disc remains the dominant component.

In Fig.~\ref{figure//chap3/ComeronDeconv} we show the thin- and thick-disc decomposition for the fit of the PSF-deconvolved model. The surface brightness of the fit reaches deeper values than for the rest of the sample ($\gtrsim$28~mag~arcsec$^{-2}$) because the profiles are less noisy and better sampled (the spatial resolution of NGC~1032 is better because of its larger apparent size). Fig.~\ref{figure//chap3/ComeronDeconv} shows that the PSF effect is almost negligible, in agreement with the 2D fitting, because the PSF-deconvolved model is systematically immediately below the observed data and has the same shape.

The  weighted mean values for the physical parameters are shown in Table \ref{table//chap3/sect3:pysParam}. In this case, the scale heights of the thick discs are the same (within errors) in the observed data as in the PSF-deconvolved model. Thus, the mass ratios are also very similar (the small difference comes from the change in the scale height of the thin disc).

\section{Discussion} \label{chap3/sec9:Discussion}

We have presented a new method as applied to a fairly diverse sample of galaxies (from early to late type ) to reach two main goals: (1) properly account for the PSF effect in different types of galaxies, and (2) parametrise the galaxy disc components by considering their physical interactions. To do this, we combined two techniques that have already been partly used in previous works (see details and references in Sect.~\ref{Fittings}). Our resulting method allowed us to reach a surface brightness limit of $\mu_{r_{\rm deep}} \sim$28.5$-$29~mag~arcsec$^{-2}$ (3\,$\sigma$, 10$\times$10\,arcsec$^{2}$). In this section we discuss the advantages and disadvantages of the different methods when the PSF effect in our sample of galaxies is analysed. We also consider possible formation and evolution scenarios of the thick disc. Throughout this discussion we use the weighted mean values of all the physical parameters of the thin and thick discs of each galaxy, as in Table~\ref{table//chap3/sect3:pysParam}.

\subsection{PSF effect} \label{chap3/sec9sub2:PSFeffect}

\cite{Sandin2014,Sandin2015} found that the scattered light of extended sources (if not accounted for) might be misinterpreted as a bright stellar halo or a thick disc. Other works have shown that the properties of stellar haloes and the faint outskirts of a galactic discs can indeed be influenced significantly by the wings of the stellar PSFs \citep[e.g.][]{Zibetti2004a, deJong2008, Sandin2014, TrujilloFliri2016, Peters2017, Comeron2017}. In general, the effect becomes stronger when the light distribution of the source is sharper \citep{Trujillo2001a, Trujillo2001b}. Here, we have used a very extended PSF for each galaxy (Sect.~\ref{chap3/sect2sub1:PSFextendedStripe 82}) in order to properly account for time, angle, and field dependencies, and with a diameter of at least twice that of the source \citep{Sandin2014}.

We used \textsc{imfit} to perform reliable 2D fitting of each galaxy and its components, producing PSF-convolved and -deconvolved models. From the surface brightness profiles in both radial and vertical directions of all the data and the models, we can quantify the PSF effects. We find that the observed radial surface brightness profile deviates less from the PSF-deconvolved profile than the vertical one, as expected. UGC~01040 is a clear example of a galaxy with a very steep distribution of light, while NGC~1032, in contrast, has very smooth surface brightness profiles along both axes.

In the radial direction, the PSF influence becomes significant  at $\mu_{r_{\rm deep}} \sim$\,28~mag~arsec$^{-2}$ (i.e. the radial surface brightness profile of the observed data starts to deviate from the PSF-deconvolved profile). At fainter levels, the scattered light contribution is up to $\sim$1.5$-$2.5 times that of the intrinsic stellar disc brightness, which is slightly lower than the values reported by \citet[][they found mass excesses up to three times higher, probably because they went much deeper]{TrujilloFliri2016}. However, it is important to note that these values were extracted from the PSF-deconvolved models without considering the residuals (i.e. all non-symmetric features such as the spiral arms that the model cannot fit) and the model was obtained from an analytical function rather than by taking the physical interactions of the discs into account.

In the vertical direction, we achieved satisfactory fits to the PSF-deconvolved models, but in this case, considering the residuals and the physical interactions of the discs (gravitationally coupled vertically isothermal thin and thick discs in hydrostatic equilibrium). In our galaxy sample, the PSF effect starts to be relevant at $\mu_{r_{\rm deep}} \sim$\,27~mag~arsec$^{-2}$ for the less massive galaxies, UGC~00931 and UGC~01839, but at $\mu_{r_{\rm deep}} \sim$\,25~mag~arsec$^{-2}$ for the galaxies of intermediate to high mass, NGC~0429 and UGC~01040 (in the high-mass galaxy NGC~1032, the PSF is negligible because its light distribution is very smooth). At lower values, the PSF (if not accounted for) adds artificial light to the outskirts of the galaxies, which corresponds to an excess in galaxy flux, and therefore in mass (for a given $M/L$), of a factor of $\sim$1.5$-$2 times for our two low-mass sources, and $\sim$2.5$-$4 times for our two galaxies of intermediate to high mass. This is a critical result that reveals that a careful PSF treatment of observed data is absolutely indispensable for deep imaging of extended objects.

We also analysed how the wings of the PSF impact each galaxy component. The PSF mainly spreads the light from the brightest regions (usually located in the centre) to the faintest (the outer parts). Thus, a galaxy component is more affected by the PSF wings if the light distribution is sharp. In particular, the intrinsic stellar disc brightness of NGC~0429 is mainly composed of the PSF wings deriving from its prominent bulge and also of the thin disc, rather than the thick disc. If not corrected for, this produces an incorrect quantification of the physical properties of the galaxy structures. In the case of  NGC~1032, the PSF-influenced thick disc is the dominant component, although the thin disc and the bulge contribute more when high altitudes above or  below the galaxy mid-plane are reached, beyond $z\sim$35~kpc \citep[see all the plots and more details of NGC~0429 and NGC~1032 in ][]{LombillaPhD}. Finally, the low surface brightness regions of UGC~01040 (top panel in Fig.~\ref{figure//chap3/UGC01040:RadVertSBP}) are dominated by the effect of the PSF wings in the thick disc, closely followed by those in the bar and the thick disc. This is important because properly accounting for the PSF is not only necessary for a better estimation of the actual sizes and masses of the galaxies, but also for an accurate determination of the shapes, sizes, and masses of their structural components.

\subsection{Possible contribution of a halo component} \label{chap3/sec9sub2:PSFeffect}

Our work reaches surface brightnesses of $\mu_{r_{\rm deep}} \sim$28.5$-$29~mag~arcsec$^{-2}$ (3\,$\sigma$, 10$\times$10\,arcsec$^{2}$) such that the stellar halo may start to become visible \citep[][and references therein]{Martin-Navarro2014}. Thus, it is worthwhile to study the cases of  NGC~0429 and NGC~1032 because in both galaxies the fit reaches regions at $R>$30~kpc. In NGC~0429, the presence of a stellar halo is not clear, as the model fits all the data points up to the very outskirts of the galaxy. However, in NGC~1032, it is not possible to obtain a good fit when all the data points in the outer parts are included. Perhaps a stellar halo component plus a thick disc with a shorter scale height (and probably also a shorter scale length) might constitute a more realistic model of the galaxy. The physical parameters of a stellar halo component are different from those of the thick disc, but it is not practical to account for a halo in this work. A logical next step for future improvement of our method therefore is to add this component into our fits. This implies accounting for physical interactions between the halo and the disc components, so that more computer time and capability will be needed as the number of free parameters will increase. Ideally, deeper imaging should also be used.

\subsection{Origin of thick discs} \label{chap3/sec9sub3:OrigTD} 

Most if not all disc galaxies have a thin (classical) disc and a thick disc. In most models, thick discs are thought to be a necessary consequence of the disc formation process and/or of the evolution of the galaxy. However, because of their faint surface brightness ($\mu_{rm g} \sim$26.0~mag~arcsec$^{-2}$), the detailed study of the physical properties of thick discs has been very difficult. We here studied their structure at significant depth by reaching a surface brightness limit of 28.5$-$29~mag~arcsec$^{-2}$ with the combined $g$, $r,$ and $i$ band images from The IAC Stripe82 Legacy Project.

\begin{figure*}
\begin{center}
\includegraphics[width=150mm]{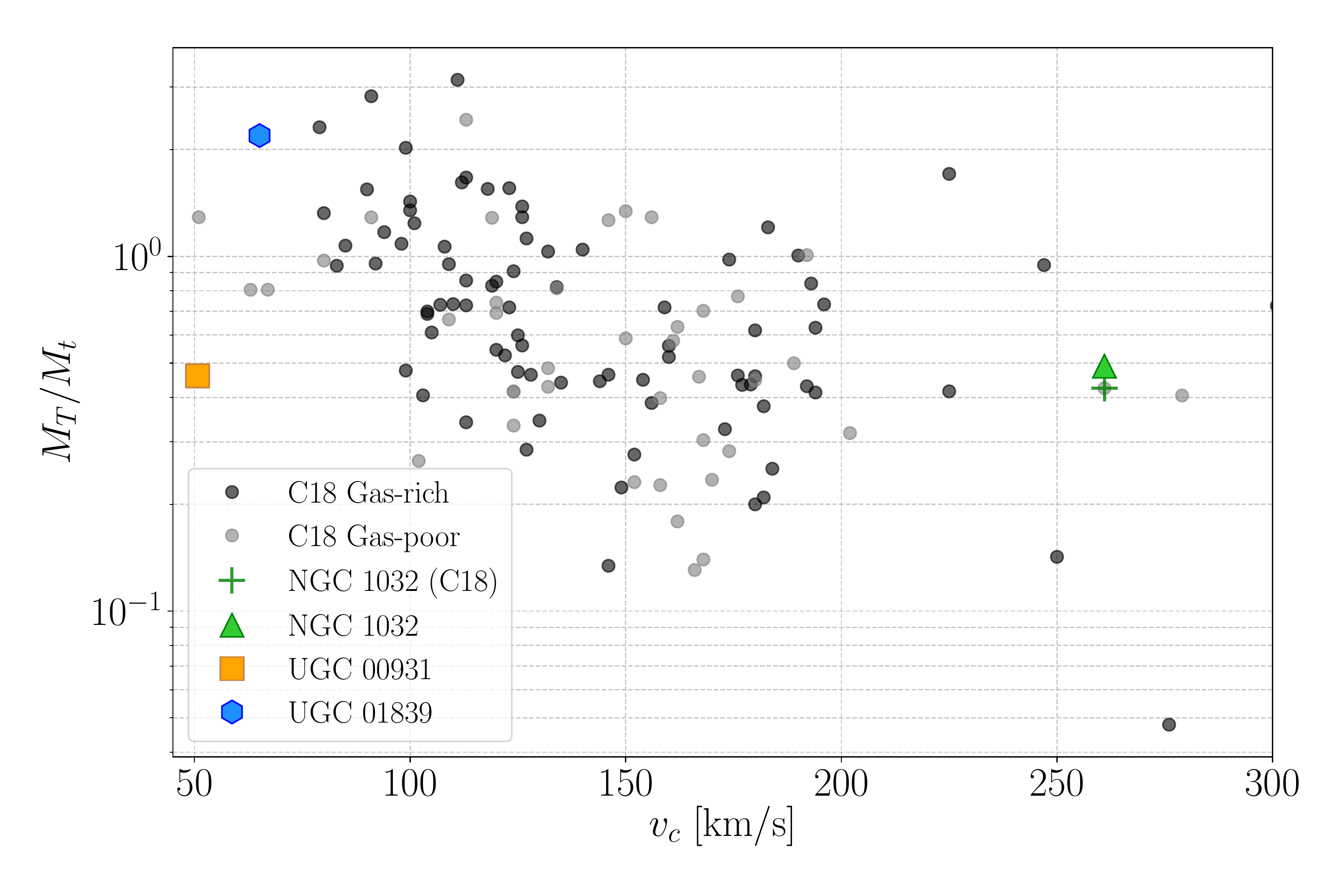} 
\protect\caption[]{Ratio of the thick- and thin-disc masses (i.e., $\Sigma_{{\rm T}}/\Sigma_{{\rm t}}$) of the PSF-deconvolved model  fits as a function of the circular velocity ($v_{c}$) for the galaxies in our sample with circular velocity information (see Table~\ref{table//chap3/sect3:finalSample}). Overplotted are the values of the sample in \citet[][]{Comeron2017}, labeled as C18.}
\label{figure//chap3/vc_MTMt}
\end{center}
\end{figure*}

\cite{Yoachim2006} and \cite{Comeron2011,Comeron2012, Comeron2017} established that thick discs are more dominant in low-mass galaxies than they are in massive galaxies. The ratio between the thick- and thin-disc masses is typically in the range 0.2~$< M_{\rm T}/M_{\rm t}<$~1 for galaxies with $v_{c} >$~120~km/s, but low-mass galaxies can have much higher values, up to $M_{\rm T}/M_{\rm t} \sim$3. In Fig.~\ref{figure//chap3/vc_MTMt} we show the thick- to thin-disc mass ratio (i.e. $\Sigma_{{\rm T}}/\Sigma_{{\rm t}}$) that results from our PSF-deconvolved model fits, plotted as a function of the circular velocity ($v_{c}$) for the galaxies in our sample with circular (i.e. rotational) velocity information (see Table~\ref{table//chap3/sect3:finalSample}). Our results are in good agreement with those of \citet[the overploted filled dots in Fig.~\ref{figure//chap3/vc_MTMt}]{Comeron2017}, and we confirm that thick discs are comparable in mass to their thin-disc counterparts.

It is common to think of a thin disc as being dominant, and embedded in a diffuse and extended thick-disc component. The low-mass galaxy characteristics point to a thick disc as the dominant component with a thinner component inside, however, which is compatible with Fig.~\ref{figure//chap3/vc_MTMt}, which shows a clear tendency towards higher $M_{\rm T}/M_{\rm t}$ when $v_{c}$ decreases (i.e. the mass of the galaxy). This in turn agrees very well with our results for the low-mass galaxies, UGC~00931 and UGC~01839, where the 2D fits suggest a single disc component \citep[see also detailed plots in Chapter~3 of][]{LombillaPhD}. In the fits to the PSF-deconvolved models of these galaxies (e.g. Fig.~ \ref{figure//chap3/ComeronDeconv}), we found two discs in both cases. However, in UGC~00931 the discs are quite similar, consistent with a single disc, whereas for UGC~01839 the thick disc clearly dominates the thin disc. This is a natural consequence in a model where thick discs were first formed from a turbulent gas disc with intense star formation \citep{Elmegreen2006, Bournaud2009, Comeron2014,Comeron2017}. This corresponds to a Universe that evolves in line with the philosophy of down-sizing, where low-mass galaxies take longer to evolve \citep[][]{Cowie1996}. In this picture, thin discs in low-mass galaxies are still in their infancy, whereas those in high-mass galaxies are already mature. An alternative interpretation may be that the star formation in the thin discs of low-mass galaxies is not efficient enough, and the thin disc cannot be distinguished from the thick disc (or could even be considered as the same structure). Thus, thick (or thin) discs might not be ubiquitous for all types of disc galaxies \citep[contrary to the result of][]{Comeron2017}.

\begin{figure*}
\begin{center}
\includegraphics[width=150mm]{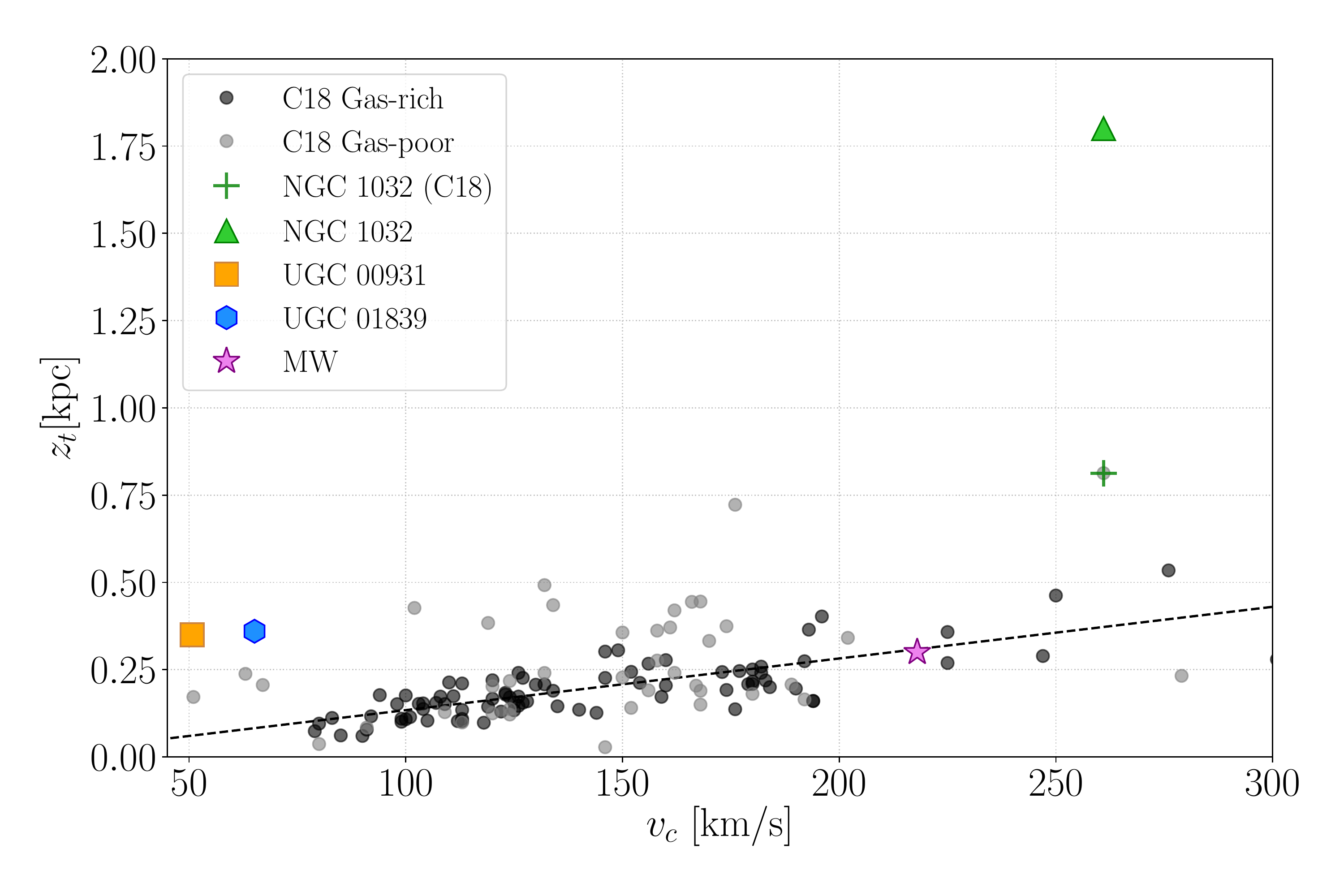} 
\includegraphics[width=150mm]{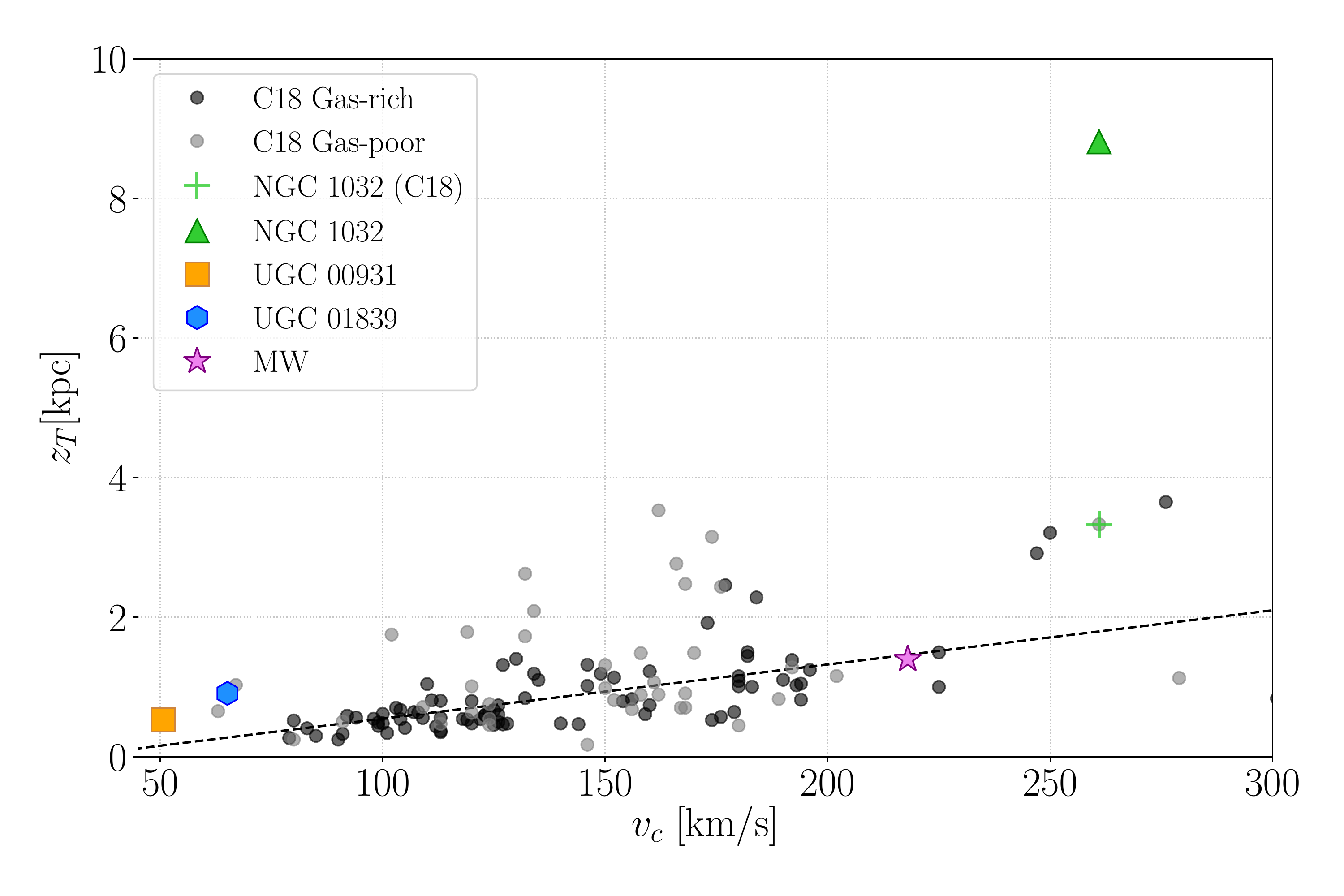}
\protect\caption[]{Thin- and thick-disc scale-heights (top and bottom panels, respectively) of the PSF-deconvolved model fit as a function of the circular velocity ($v_{c}$) for the galaxies in our sample with circular velocity information (see Table~\ref{table//chap3/sect3:finalSample}). The stars represent the corresponding values for the Milky Way (scale heights from \citealt{Gilmore1983}, and circular velocity from \citealt{Bovy2012a}). The black dashed line corresponds to a robust linear regression to the gas-rich galaxies in \cite{Comeron2017}. Overplotted are the values of \cite{Comeron2017}, labelled C18.}
\label{figure//chap3/vc_zt}
\end{center}
\end{figure*}

Figure~\ref{figure//chap3/vc_zt} shows the scale heights of the thin ($z_{{\rm t}}$) and the thick ($z_{{\rm T}}$) discs as a function of the circular velocity ($v_{c}$) for the three galaxies in the sample with available circular velocity values \citep[equivalent to Figure 17 from][]{Comeron2017}. Compared to the results of \citet{Comeron2017}, UGC~00931 and UGC~01839 fall in the expected area when we consider the uncertainties in the distance measurements and all the uncertainties associated with our rather complicated fits. The thin- and thick-disc scale heights of these galaxies correlate with the circular velocities of the galaxies, in good agreement with \cite{Comeron2017}. In particular, the positions of UGC~00931 and UGC~01839 in Fig.~\ref{figure//chap3/vc_zt} seem to fit the gas-poor galaxy sample better.

NGC~1032 is the only galaxy in common with the sample of \citet{Comeron2017}, so that here we show an optical (and deeper) counterpart to their work. When we return to Fig.~\ref{figure//chap3/vc_zt}, the data point for NGC~1032 is even more of an outlier than in \cite{Comeron2017} for the thin- and thick-disc scale heights. This galaxy is a gas-poor object and has one of the two largest scale heights for a thin disc in the results of \cite{Comeron2017}. However, our results suggest a higher discrepancy in the scale heights of both discs ($z_{{\rm t, C18}}=0.8$~kpc versus $z_{{\rm t, TW}}=$1.8~kpc, and $z_{{\rm T, C18}}=$3.3~kpc versus $z_{{\rm t, TW}}=$8.8~kpc; where we used C18 for \citealt{Comeron2017}, and TW for results of PSF-deconvolved models in this work), even though the scatter in the relation is larger for gas-poor galaxies than for gas-rich ones. As NGC~1032 is isolated \citep{Tully2015}, the disc heating may have been caused by the merger with a companion.

The scale heights of the thin and thick discs of NGC~1032 are very different from those reported by \cite{Comeron2017}, which might be due to various reasons. The main differences between this work and that of \citet[][]{Comeron2017} are the different wavelength ranges, the PSF treatment, and the depth of the data. The latter is a key factor as it may allow us to see the stellar halo, so that our thin- and thick-disc scale heights could be larger that they otherwise may be. In the case of NGC~1032, we already saw that the PSF does not cause significant effects because the light distribution of the galaxy is very smooth. On the other hand, each wavelength gives different information about the underlying stellar population properties. Because near-infrared light traces older populations than the optical, our larger scale heights could point to a young stellar population component in the outer parts of both discs. In agreement with this, the thick- to thin-disc scale height ratios of this galaxy are quite similar in both works:  $(z_{{\rm T}} / z_{{\rm t}})_{\rm C18}=$4.1, and $(z_{{\rm T}} / z_{{\rm t}})_{\rm TW}=$4.9. Another fact in favour of this hypothesis is that the thick- to thin-disc mass ratios are also similar:  $(\Sigma_{{\rm T}}/\Sigma_{{\rm t}})_{\rm C18}=$0.4, and $(\Sigma_{{\rm T}}/\Sigma_{{\rm t}})_{\rm TW}=$0.5. When we take the thick- to thin-disc mass ratio of our observed data ($(\Sigma_{{\rm T}}/\Sigma_{{\rm t}})_{\rm TW obs}=$0.4), the result is even more similar, which may suggest that the scattered light does not affect the ratio much.

Comparing all our fits with those in \cite{Comeron2017}, we reach deeper surface brightness levels, which highlights the PSF effect over galaxy structures beyond $\mu_{r_{\rm deep}} \sim$27~mag~arcsec$^{-2}$. While \citet[][]{Comeron2017} did not find a significant excess of mass due to scattered light, we obtain overestimates in the mass of the thick disc of up to a factor of four in the PSF-deconvolved model than in the data at $\sim$28~mag~arcsec$^{-2}$. However, \cite{Comeron2017} found that even though the thin disc is less extended vertically than the thick disc, it dominates the emission at large heights because of the light scattered from the mid-plane to large heights, in agreement with our results. This effect would remain unnoticed if we had not accounted for the extended wings of the PSF. \\

We finish by analysing the mass-to-light ratios between the thin and thick discs ($(M/L)_{\rm T}/(M/L)_{\rm t} = \Upsilon_{\rm T} / \Upsilon_{\rm t}$, obtained as described in Sect.~\ref{chap3/sec6sub1sub1:M/L}), which gives an idea of the stellar population age of each component. For the 2D PSF-deconvolved model of our sample of galaxies, we obtain the following values: 0.99, 1.07, 1.97, 1.46, and 0.56 (for NGC~0429, UGC~01040, UGC~00931, UGC~01839, and NGC~1032, respectively, with $\sim$15\,\% uncertainty). In the cases of NGC~0429 and UGC~01040, $(M/L)_{\rm T}/(M/L)_{\rm t}$ is almost unity, so that both disc components seem to have similar stellar populations, suggesting a common origin for the thin- and thick-disc components. NGC~1032 shows younger stellar populations in the thick disc than those in the thin disc ($(M/L)_{\rm T}/(M/L)_{\rm t}=$0.56). In the low-mass galaxies UGC~00931 and UGC~01839, the thick disc looks older than the thin disc, in agreement with the hypothesis that their thick discs formed first from a turbulent gas disc with intense star formation \citep{Elmegreen2006, Bournaud2009, Comeron2014,Comeron2017}.

In general, our results are in good agreement with those in \cite{Comeron2017}, although there are some discrepancies because we analysed data in a different wavelength and down to a deeper surface brightness limit. This suggests that our method is able to properly reproduce the disc components beyond $\mu_{r_{\rm deep}} \sim$27~mag~arcsec$^{-2}$, bringing light to the formation and evolution scenario of low surface brightness galaxy structures.

Although our sample of galaxies is small, we find that the three formation scenarios proposed to explain the origin of the thick discs are likely. It seems that the formation of these structures depends on the galaxy mass and morphology, and on the mass assembly history of a given galaxy. In low-mass galaxies, UGC~00931 and UGC~01939, only a single disc component is required, in agreement with a model where thick discs were first formed from a turbulent gas disc with intense star formation. In the intermediate-high mass galaxies NGC~0429 and UGC~01040, a common origin is suggested for both the thin- and thick-disc components. Finally, NGC~1032 seems to have formed a thick disc as a result of merger events.

\section{Conclusions}

We characterised the thick discs of five edge-on galaxies down to a surface brightness limit of $\mu_{r{\rm deep}} \sim28.5-29$~mag~arcsec$^{-2}$ (3\,$\sigma$, 10$\times$10\,arcsec$^{2}$) with combined $g$-, $r$-, and $i$ -band images from the IAC Stripe 82 Legacy Project. The method accounts for a careful analysis of the background, masking, and a meticulous PSF treatment through 2D galaxy modelling. Our results are based on a study of radial and vertical surface brightness profiles, and we compared our data with models. The galaxy disc components were fitted by considering that the thin and thick discs are two stellar fluids in hydrostatic equilibrium. The main conclusions of this work are listed below.

\begin{enumerate}

\item  We developed and tested a new method and applied it to a fairly diverse sample of five galaxies, from early- (i.e. S0) to late-type galaxies (Sdm). The main goals were to 1) properly account for the PSF effect in very deep imaging of extended objects, and 2)  reproduce their disc components in unprecedented detail. 

\item We used this technique to reproduce 2D PSF-deconvolved galaxy models (including low surface brightness structures) and to obtain thin- and thick-disc decompositions by considering physical interactions. The method is highly time-consuming, therefore large galaxy samples cannot be analysed at present.
  
\item A careful PSF treatment of observed data is absolutely indispensable for deep imaging of extended objects. The PSF effect (if not accounted for) produces an overestimate of the mass in the galaxy outskirts at $\sim$28~mag~arcsec$^{-2}$ by a factor of 1.5--2 in our two low-mass sources and of 2.5--4 in our two intermediate- to high-mass galaxies (previous works such as \citealt{Comeron2017} did not find this mass excess because they reached $\sim 2$~mag~arcsec$^{-2}$ brighter than we did). The wings of the PSF affect the structures (not only the thick discs) by adding artificial light to the outer parts, which hinders the determination of the real shapes and sizes of galaxy components.

\item A logical next step to improve our method is to add a stellar halo component to our fits. This would account for physical interactions between the halo and the disc components, but more computer time is required because the number of free parameters will increase. 

\item Our sample of five galaxies provides evidence for aspects of a wide variety of the formation scenarios of thick discs in disc galaxies. The origin of these structures depends on the galaxy mass and morphology, and on the mass assembly history of a given galaxy:

\begin{enumerate}
\item In 2D fittings of the low-mass galaxies UGC~00931 and UGC~01839, only a single disc component is required. The mass-to-light ratios between the thin and thick discs is $(M/L)_{\rm T}/(M/L)_{\rm t}>$1 for these galaxies, suggesting a redder and older thick disc. These two arguments are in agreement with a model where thick discs were first formed from a turbulent gas disc with intense star formation \citep{Elmegreen2006, Bournaud2009, Comeron2014,Comeron2017}. An alternative interpretation is a thin disc with low star formation efficiency, which cannot be distinguished from the thick disc (it could even be considered as the same structure). Thus, thick (or thin) discs might not be ubiquitous for all types of disc galaxies \citep[contrary to the result of][]{Comeron2017}. 

\item In general, our results on scale heights and mass ratios of thin and thick discs ($z_{{\rm t}}$, $z_{{\rm T}}$, $M_{{\rm T}}/M_{{\rm t}}$, respectively) confirm the results of \cite{Comeron2017} (see Figs \ref{figure//chap3/vc_MTMt} and \ref{figure//chap3/vc_zt}). In the case of the one galaxy we have in common, NGC~1032, the main discrepancy is in the values of $z_{{\rm t}}$ and $z_{{\rm T}}$, which are perhaps due to a young stellar population component that the near-infrared data cannot trace.

\item For NGC~0429 and UGC~01839, $(M/L)_{\rm T}/(M/L)_{\rm t}$ is almost unity, so that both disc components seem to have similar stellar populations. This suggests a common origin for the thin- and thick-disc components.

\end{enumerate}

\end{enumerate}

\begin{acknowledgements}
     We thank Ignacio Trujillo, Sebasti{\'e}n Comer{\'o}n, Victor Debattista, and Lee Kelvin for many useful suggestions and comments. We acknowledge support from the Spanish Ministry of Economy and Competitiveness (MINECO) under grant numbers AYA2013-41243-P and AYA2016-76219-P. J.H.K. acknowledges financial support from the European Union's Horizon 2020 research and innovation programme under Marie Sk$\l$odowska-Curie grant agreement No. 721463 to the SUNDIAL ITN network, from the Fundaci{\'o}n BBVA under its 2017 programme of assistance to scientific research groups, for the project ``Using machine-learning techniques to drag galaxies from the noise in deep imaging", and from the Leverhulme Trust for the award of a Visiting Professorship at Liverpool John Moores University. CML thanks the Astrophysics Research Institute of Liverpool John Moores University for their hospitality, and the Spanish Ministry of Science, Innovation, and Universities for financial support of her visit there, through grant number EST2019-013110-I. This research has made use of different Python packages and of NASA's Astrophysics Data (NED) System Bibliographic Services which is operated by the Jet Propulsion Laboratory, California Institute of Technology, under contract with the National Aeronautic and Space Administration. 
\end{acknowledgements}

%-------------------------------------------------------------------

%%%%%%%%%%%%%%%%%%%%%%%%%%%%%%%%%%%%%%%%%%%%%%%%%%



\bibliographystyle{aa}

%%%%%%%%%%%%%%%%%%%%%%%%%%%%%%%%%%%%%
%%%%%%%%%%%%%%%%%%%%%%%%%%%%%%%%%%%%%

%%%%%%%%%%%%%%%%%%%%%%%%%%%%%%%%%%%%%
%%%%%%%%%%%%%%%%%%%%%%%%%%%%%%%%%%%%%

\begin{appendix}  %First appendix

\section{Bootstrap resampling of the 2D galaxy models} \label{appendix}

Each 2D model of the galaxies in our sample was obtained using the astronomical software \textsc{imfit} \citep{Erwin2015}. However, as described in Sect. \ref{Sect:modFitting}, each galaxy is made of components that need an analytical function that describes its light distribution. To model just one component, it is necessary to leave at least three parameters free. In the case of two or more functions, the image fits were first performed for the innermost components (i.e. bulge and/or bar) and then, with the parameters for those functions fixed, the parameters of the components in the outer part (i.e. thin and/or thick disc) were left free. The output for a reasonable fit from such an initial run was then used to re-run the code by allowing all (or most) model parameters to be free. However, there is a strong degeneracy of the disc parameters, which can result in multiple possible solutions for the modelling process: two or more different solutions can constitute a similarly good fit for a given galaxy. In order to solve this degeneracy problem, \textsc{imfit} allows bootstrap resampling. Bootstrap resampling is a method for estimating the variability of our results by making a more detailed analysis of the parameter distributions, including potential correlations between free parameters. The combined set of bootstrapped parameter values can be used to estimate confidence intervals, and in consequence, to discern between the multiple good solutions. With this procedure, we were able to ensure that the 2D PSF-convolved model fits in our work are not degenerate.

In the following figures we show a set of bootstrapped parameter values and their confidence intervals obtained from the 2D model fitting of each edge-on galaxy in our sample. The bootstrap resampling performed 1500 iterations of each fitting process. In general, the parameters for the discs were the central intensity ($J_{0}$), the value of which for each pixel comes from line-of-sight integration through a 3D luminosity-density model, generating a projected 2D model image given input specifications of the orientation and inclination to the line of sight (i.e. it is not the counts in the image, therefore is not straightforward to convert into surface brightness units); the inner and outer exponential scale lengths ($h_{1}$, $h_{2}$) in kpc for broken exponential functions; the exponential scale lengths ($h$) in kpc for exponential functions; and the vertical scale heights ($z$) also in kpc. When the disc has distinct thin and thick disc components, the corresponding values are indicated with the subscripts $t$ and $T$, respectively. If there is just one disc component, it is shown with the subscript~$d$.

\begin{figure*}
\begin{center}
\includegraphics[width=180mm]{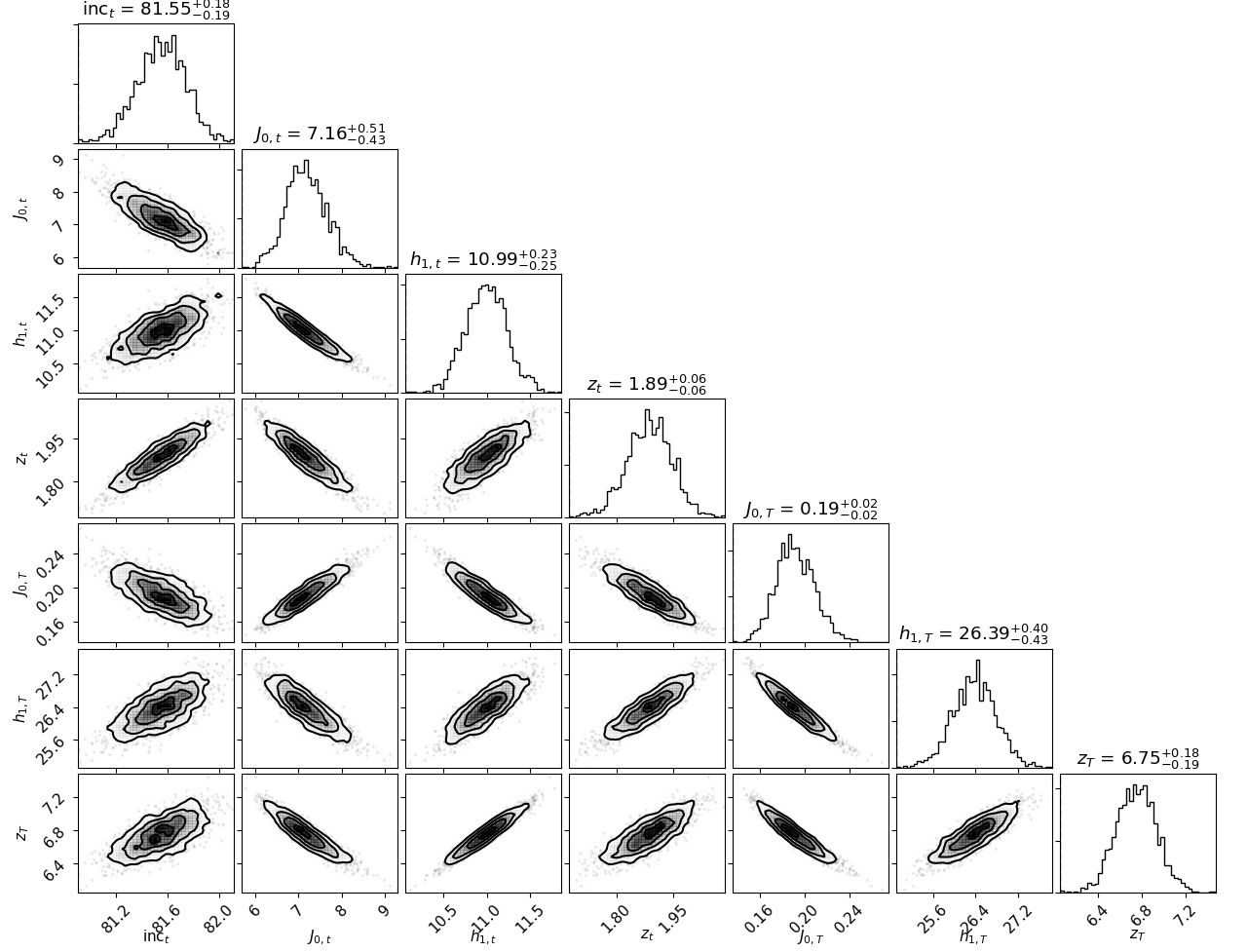}
\protect\caption[ ]{Set of bootstrapped parameter values and their confidence intervals obtained from the 2D model fitting of the edge-on galaxy NGC~0429. In this case, we also checked the inclination in degrees between the line of sight and the polar axis of the thin disc (${\rm inc}_{{\rm thin}}$) because it was an uncertain parameter for this galaxy. }
\label{figure//chap3/NGC0429:bootstrap} 
\end{center}
\end{figure*}

\begin{figure*}
\begin{center}
\includegraphics[width=180mm]{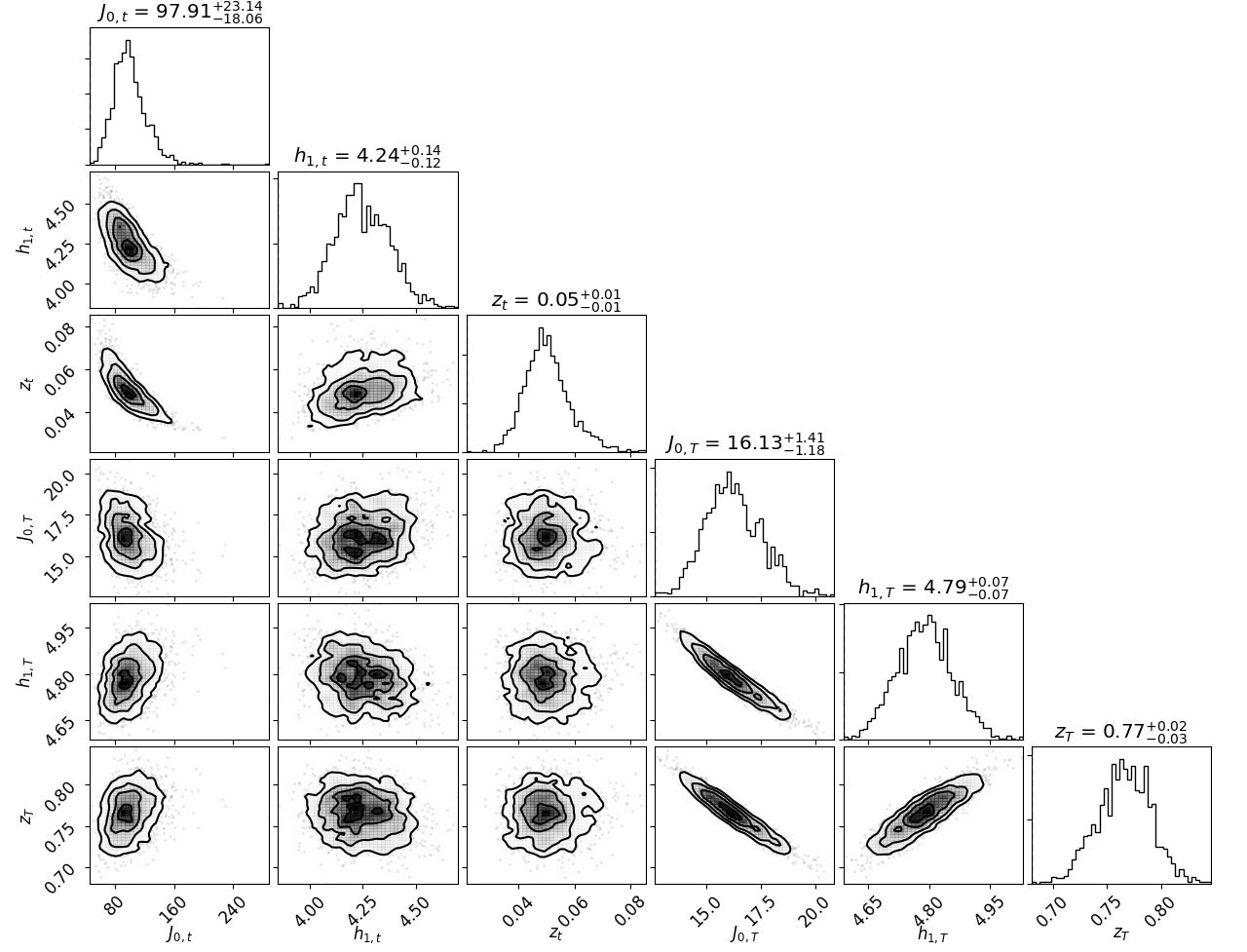}
\protect\caption[ ]{Same as Fig. \ref{figure//chap3/NGC0429:bootstrap}, now for the edge-on galaxy UGC~1040.}
\label{figure//chap3/UGC01040:bootstrap} 
\end{center}
\end{figure*}

\begin{figure*}
\begin{center}
\includegraphics[width=.55\linewidth]{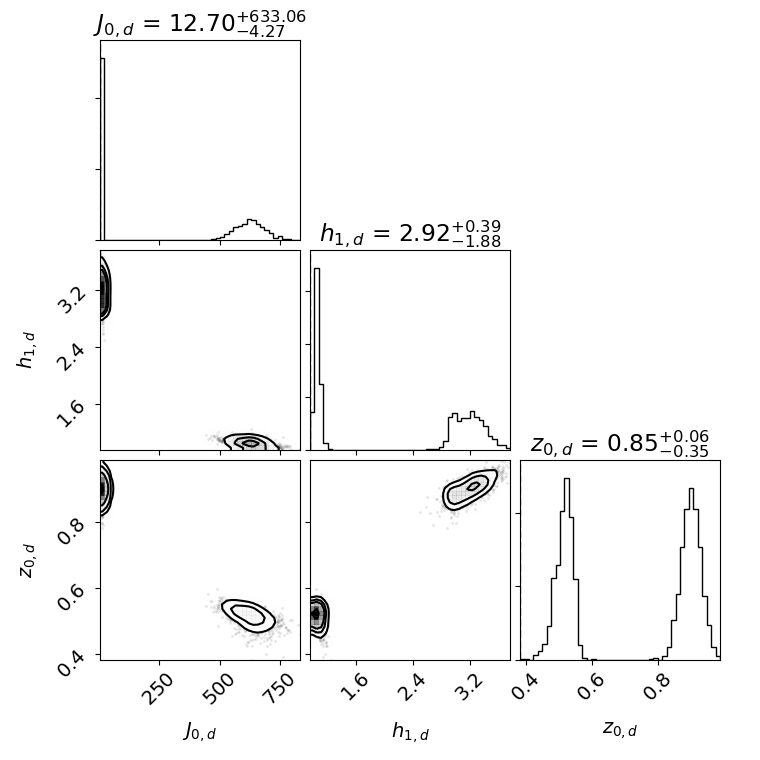}
\includegraphics[width=.55\linewidth]{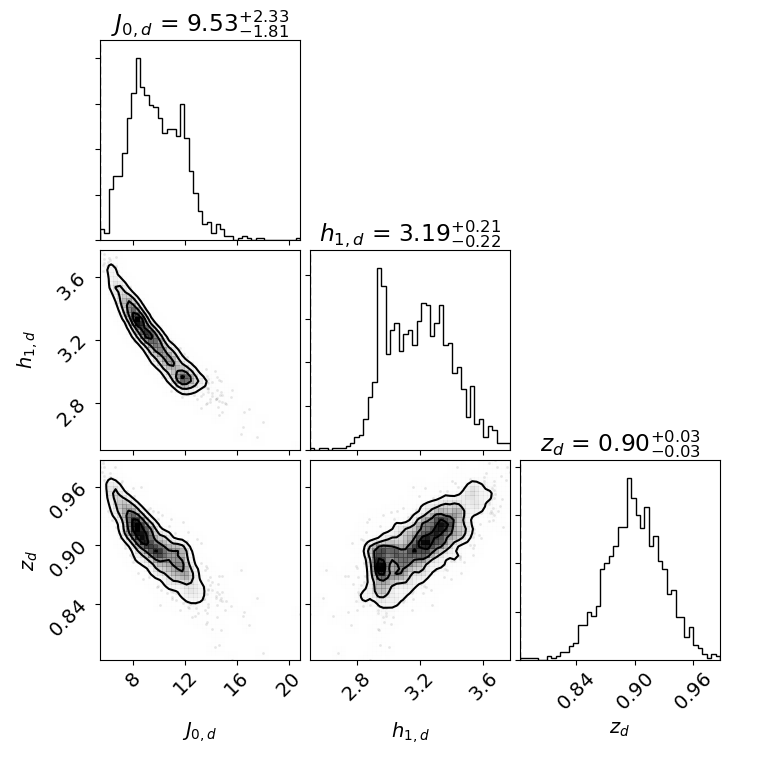}
\protect\caption[ ]{Same as Fig. \ref{figure//chap3/NGC0429:bootstrap}, now for the edge-on galaxy UGC~00931. However, in this galaxy we had a particular situation: we obtained two possible solutions for the fit that are shown in the two panels (upper panel 2500 iterations, lower panel 1500 iterations). We chose the set of parameters that was most likely and had greater physical sense, regarding the physical parameters values and the surface brightness profiles shape. Thus, we selected the option of the bottom panel. In the upper left panel in the panel at the top and in the corresponding profile generated by this model (Fig.~\ref{figure//chap3/UGC00931:badSBP}), we see that the main reasons for discrepancy between both models are in the central intensity of the disc ($J_{0, d}$) as $J_{0,d}=12.7^{+633.06}_{-4.27}$~values/pix, and in the light distribution (shown in profile of Fig.~\ref{figure//chap3/UGC00931:badSBP}). Thus, the set of chosen values is $J_{0,d}=9.53^{+2.33}_{-1.81}$~values/pixel, $h_{1, d}=3.19^{+0.21}_{-0.22}$~kpc, and $z_{d}=0.90^{+0.03}_{-0.03}$~kpc (lower panel).}
\label{figure//chap3/UGC00931:bootstrap} 
\end{center}
\end{figure*}

\begin{figure*}
\begin{center}
\includegraphics[width=160mm]{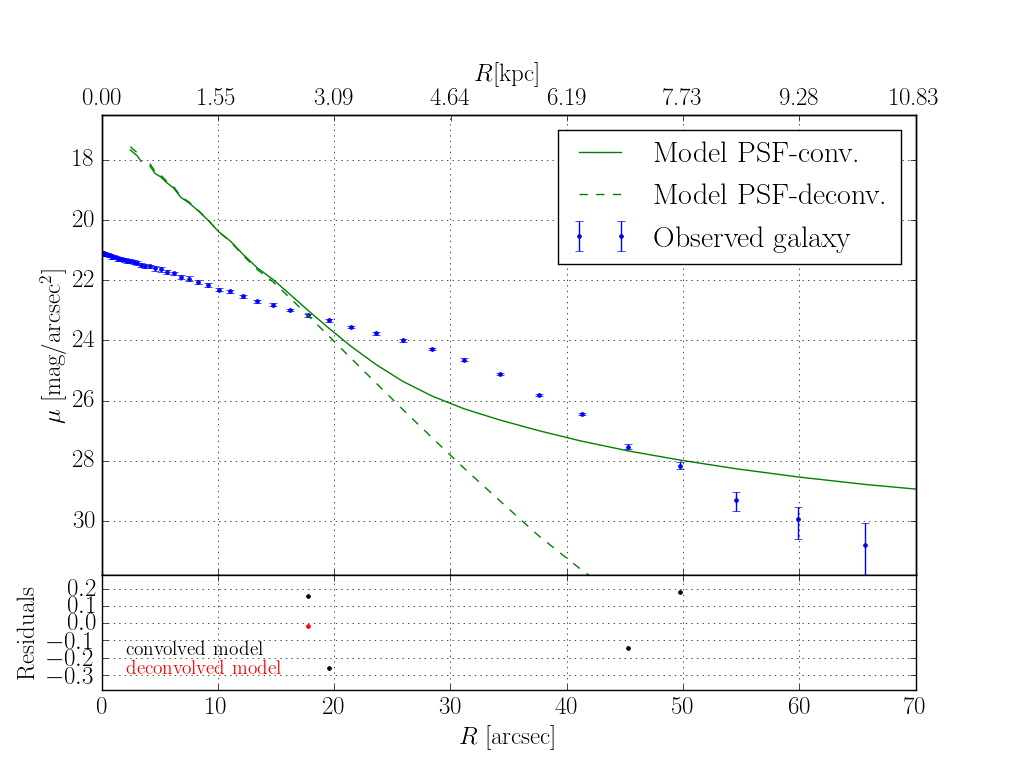}
\protect\caption[ ]{Radial surface brightness profile along the mid-plane of the edge-on galaxy UGC~00931 in the combined $r_{\rm deep}$ band of the IAC Stripe 82 Legacy Project data. We plot our data with blue dots, the PSF-convolved model is shown with the solid green curve, and the PSF-deconvolved model is the dashed curve. In the lower panel, we show the difference between the observed data and either the PSF-convolved model (black points) or the PSF-deconvolved models (red points). Here we show the result for the rejected set of parameters when the galaxy is modelled (top panel in Fig.~\ref{figure//chap3/UGC00931:bootstrap}). The 2D model clearly does not fit the data.}
\label{figure//chap3/UGC00931:badSBP} 
\end{center}
\end{figure*}

\begin{figure*}
\begin{center}
\includegraphics[width=130mm]{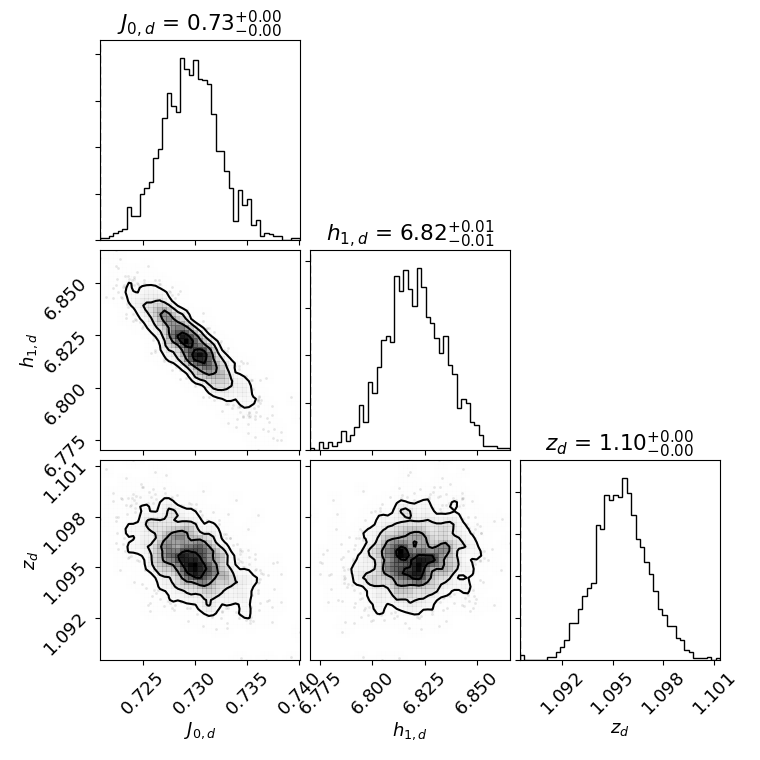}
\protect\caption[ ]{Same as Fig. \ref{figure//chap3/NGC0429:bootstrap}, now for the edge-on galaxy UGC~01839. }
\label{figure//chap3/UGC01839:bootstrap} 
\end{center}
\end{figure*}

\begin{figure*}
\begin{center}
\includegraphics[width=180mm]{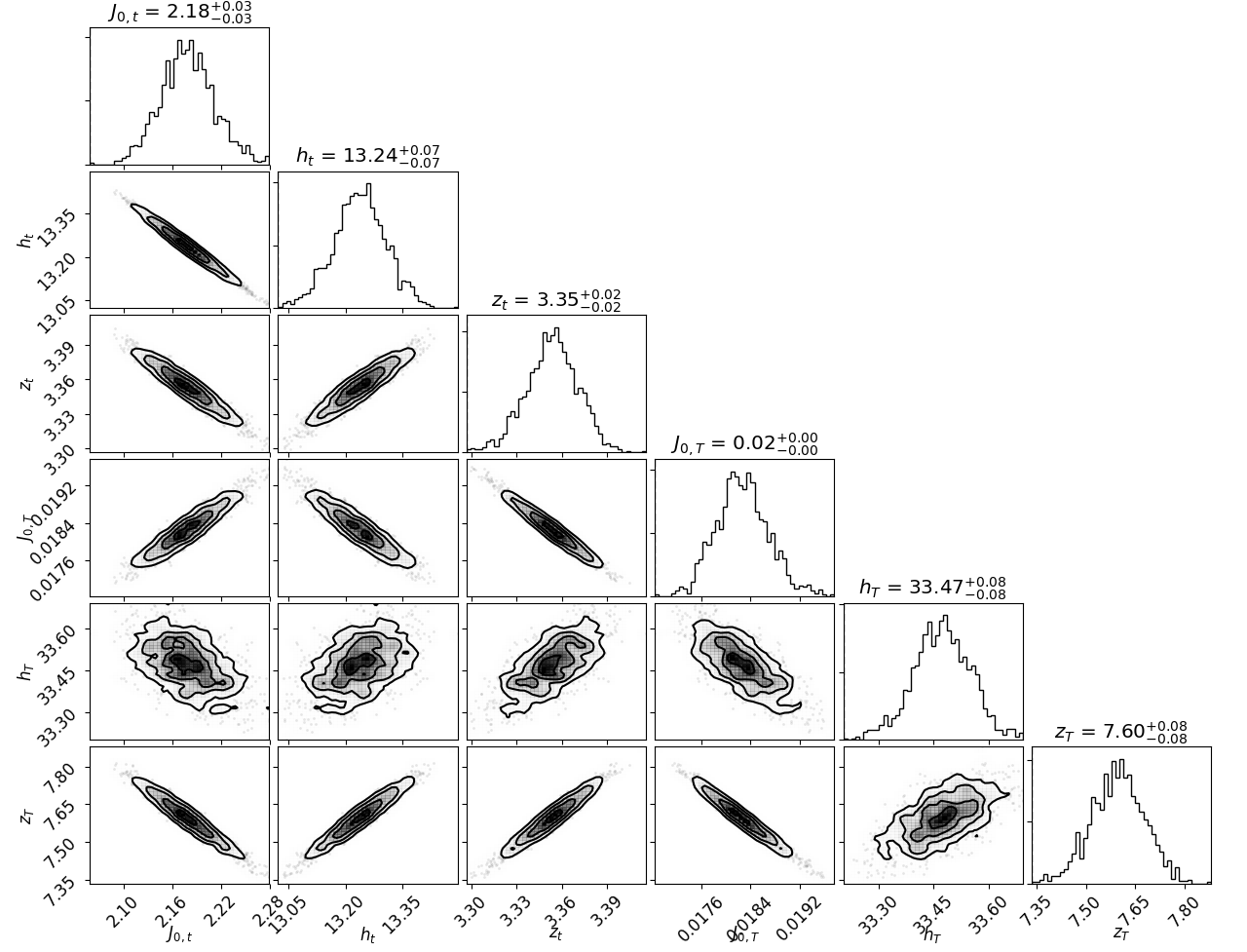}
\protect\caption[ ]{Same as Fig. \ref{figure//chap3/NGC0429:bootstrap}, now for the edge-on galaxy NGC~1032.}
\label{figure//chap3/NGC1032:bootstrap} 
\end{center}
\end{figure*}

Figures \ref{figure//chap3/NGC0429:bootstrap}, \ref{figure//chap3/UGC01040:bootstrap}, \ref{figure//chap3/UGC01839:bootstrap}, and \ref{figure//chap3/NGC1032:bootstrap} show the results for NGC~0429, UGC~01040, UGC~01839, and NGC~1032, respectively. For all these galaxies, one single set of parameters is obtained as the only (and not degenerate) possible solution. However, for UGC~00931 (Fig.~\ref{figure//chap3/UGC00931:bootstrap}), we obtain two different solutions, mainly due to a high discrepancy in the central intensity ($J_{0, d}$) value. We chose the set of parameters that is most physically meaningful and better fits the surface brightness profiles (see Fig.~\ref{figure//chap3/UGC00931:badSBP}). This is the set with lower central intensity, $J_{0,d}=9.53^{+2.33}_{-1.81}$~values/pixel, $h_{1, d}=3.19^{+0.21}_{-0.22}$~kpc, and $z_{d}=0.90^{+0.03}_{-0.03}$~kpc.\\

\end{appendix}

\end{document}